\documentclass[twocolumn,twocolappendix]{aastex63}

\usepackage{mathrsfs}
\usepackage{mathtools}
\usepackage[retainorgcmds]{IEEEtrantools}
\usepackage{graphics}
\usepackage{graphicx}
\usepackage{xcolor}
\usepackage{tabularx}
\usepackage{enumerate}

\newcommand{\ud}{\mathrm{d}}

\defcitealias{2020A&A...641A...6P}{Planck 2018 results VI. Cosmological parameters}

\shorttitle{21~cm Linear Polarization from the EoR}
\shortauthors{Li, Tan \& Mao}

\begin{document}

\title{Linear Polarization of the 21 cm Line from the Epoch of Reionization}

\correspondingauthor{Bohua Li, Yi Mao}
\email{bohuali@tsinghua.edu.cn (BL), ymao@tsinghua.edu.cn (YM)}

\author[0000-0002-3600-0358]{Bohua Li}
\affiliation{Department of Astronomy,
Tsinghua University,
Beijing, 100084, China}

\author[0000-0001-6161-7037]{Jianrong Tan}
\affiliation{Department of Astronomy,
Tsinghua University,
Beijing, 100084, China}
\affiliation{Department of Physics \& Astronomy, University of Pennsylvania, 209 South 33rd Street, Philadelphia, PA 19104, USA}

\author[0000-0002-1301-3893]{Yi Mao}
\affiliation{Department of Astronomy,
Tsinghua University,
Beijing, 100084, China}




\begin{abstract}

The 21~cm linear polarization due to Thomson scattering off free electrons 
can probe the distribution of neutral hydrogen in the intergalactic medium during the epoch of reionization,
complementary to the 21~cm temperature fluctuations.
Previous study \citep{2005ApJ...635....1B} estimated the strength of polarization with a toy model and claimed that it can be detected with 1-month observation of the Square Kilometre Array (SKA). 
Here we revisit this investigation with account of nonlinear terms due to inhomogeneous reionization, using seminumerical reionization simulations to provide the realistic estimation of the 21~cm TE and EE angular power spectra ($C^{\rm TE}_\ell$ and $C^{\rm EE}_\ell$). 
We find that (1) both power spectra are enhanced on sub-bubble scales but suppressed on super-bubble scales, compared with previous results; 
(2) $C^{\rm TE}_\ell$ displays a zero-crossing at $\ell<100$, and its angular scale is sensitive to the scale-dependence of H~{\small I} bias on large scales; 
(3) the ratios of the power spectrum to its maximum value during reionization at a given $\ell$, i.e.\ $C^{\rm TE}_\ell / C^{\rm TE}_{\ell,{\rm max}} $ and $C^{\rm EE}_{\ell}/C^{\rm EE}_{\ell,{\rm max}}$, show robust correlations with the global ionized fraction. 
However, measurement of this signal will be very challenging not only because the overall strength is weaker than the sensitivity of SKA, but also because of
the polarized foregrounds from diffuse synchrotron emission, and Faraday rotation which modifies the observed polarization. 
Nevertheless, the 21~cm linear polarization signal may still likely be detectable through other approaches, e.g.\ its cross-correlation with other probes.

\end{abstract}

\keywords{Reionization (1383), H~{\small I} line emission (690), Radio interferometry (1346), Intergalactic medium (813), Large-scale structure of the universe (902), Two-point correlation function (1951)}


\section{Introduction} \label{sec:intro}

The epoch of reionization (EoR) is a major phase transition of the Universe,
during which the neutral hydrogen in the intergalactic medium (IGM)
is heated and ionized by ultraviolet and X-ray photons from the first luminous objects. 
The process of cosmic reionization involves the formation of the large-scale structure and the rich astrophysics associated with the formation of first luminous objects.
Current constraints on the EoR \citep{2015ApJ...811..140B,2015ApJ...802L..19R,2019ApJ...879...36F} are from robust but indirect probes, e.g., the observations of high-redshift quasar spectra \citep[e.g.,][]{2006AJ....132..117F, 2015MNRAS.447..499M, 2015MNRAS.447.3402B}, and the electron scattering optical depth to the cosmic microwave background (CMB) 
\citepalias{2020A&A...641A...6P}, in the sense that these probes are only sensitive to some, but not all, information of the EoR. 

The most promising direct avenue of probing the EoR
is the intensity mapping of the redshifted 21~cm line due to the hyperfine transition of atomic hydrogen \citep[e.g.,][]{2012RPPh...75h6901P}, because the 21~cm tomography can map the distribution of neutral hydrogen (H~{\small I}) for a broad range of redshifts along the light cone. This tomographic mapping, which reveals the global history and morphological structure of cosmic reionization, contains a wealth of information regarding the structure formation of the Universe \citep[e.g.,][]{1990MNRAS.247..510S, 2006ApJ...653..815M, 2008PhRvD..78b3529M, 2019ApJ...885...23C}, 
and the properties of the first galaxies and quasars as the source of reionization 
\citep[e.g.,][]{2010A&A...523A...4B, 2015ApJ...802....8A, 2016MNRAS.456.3011D}.

The next decades will be a golden age for the 21~cm observations. 
Current interferometric arrays, e.g.\ the Murchison Wide field Array (MWA, \citealp{2013PASA...30....7T}), 
the LOw Frequency Array (LOFAR, \citealp{2013A&A...556A...2V}), 
the Precision Array for Probing the Epoch of Reionization (PAPER, \citealp{2010AJ....139.1468P}), 
and the Giant Metrewave Radio Telescope (GMRT, \citealp{2017A&A...598A..78I}), 
have first attempted to put upper limits on the 21~cm power spectrum from the EoR \citep{2013MNRAS.433..639P,2015ApJ...809...62P,2020MNRAS.493.1662M,2020MNRAS.493.4711T}. 
The next-generation radio interferometer arrays, 
including the Hydrogen Epoch of Reionization Array (HERA, \citealp{2017PASP..129d5001D}) 
and the Square Kilometre Array (SKA, \citealp{2013ExA....36..235M,2015aska.confE...1K}), 
promise to measure the statistical fluctuations of the 21~cm signal from the EoR for the first time. 
Furthermore, the SKA will very likely have enough sensitivity to generate the tomographic 21~cm maps.



The major efforts in the 21~cm modeling and data analysis \citep[e.g.,][]{1997ApJ...475..429M,2004ApJ...613...16F, 2004ApJ...615....7M, 2008ApJ...680..962L, 2019MNRAS.487.3050H, 2020PASP..132f2001L} have hitherto focused on the total intensity (i.e.\ Stokes parameter $I$) of the 21~cm brightness temperature. 
However, the polarization of the 21~cm radiation field (i.e.\ Stokes parameter $Q$, $U$ and $V$), 
as a complementary probe to the 21~cm temperature field, also contains independent information of the EoR. 
In fact, radio interferometers, as constructed by pairs of orthogonal dipole antennae, 
can be sensitive to the polarization information after careful calibration \citep{2020PASP..132f2001L}. 
Thorough investigations of the 21~cm polarization signal, therefore, 
are worthwhile in order to obtain additional science returns from the interferometric observations. 
This paper is dedicated to the 21~cm {\it linear} polarization (i.e.\ Stokes parameter $Q$ and $U$) from the EoR. 
Note that the circular polarization (i.e.\ Stokes parameter $V$) of the 21~cm signal 
can also be produced through some mechanisms, 
e.g.\ the Zeeman splitting of the triplet state by local magnetic fields \citep{2005MNRAS.359L..47C}, 
and the splitting induced by the CMB quadrupole during the dark ages \citep{2018PhRvD..97j3521H, 2021PhRvD.103b3516J}. However, the 21~cm circular polarization signal is subdominant during the EoR, and therefore not the focus of this paper.


\citet[][hereafter BL05]{2005ApJ...635....1B} pioneered the study of the 21~cm linear polarization signal, 
and discussed two categories of production mechanisms ---  intrinsic and secondary mechanisms. 
In principle, anisotropic Ly$\alpha$ pumping can produce a small signal of linear polarization intrinsically, 
i.e.\ at the time of emission of the 21~cm signal. 
The dominant effect, however, is due to the Thomson scattering of 21~cm photons off free electrons during the EoR. 
In analogy with the CMB polarization, the 21~cm radiation, even if unpolarized at the time of emission, 
will become partially linearly polarized when the scattering electrons see a quadrupole anisotropy in the temperature fluctuations, 
so this effect was called a secondary mechanism. 
In particular, for scalar perturbations, Thomson scattering only induces $E$-mode polarization due to symmetry argument. 

BL05 studied the anisotropy of the 21~cm $E$-mode polarization 
and derived the EE and TE angular power spectra. 
As the first study of this subject, 
BL05 calculated the first-order components in the power spectrum, 
and employed a simple ansatz for the ionization power spectrum. 
On the observational side, BL05 forecast that the 21~cm TE power spectrum 
could be detected by the SKA with 1-month integration time, which  
might be too optimistic.

In this paper, we will revisit the formalism for the 21~cm linear polarization from the EoR, 
taking into account the nonlinear effects due to inhomogeneous reionization. 
These include (1) the coupling between the density fluctuations 
and the nonlinear ionization fraction fluctuations, 
and (2) the cross-power between the nonlinear H~{\small I} field 
and the peculiar velocity field due to the redshift-space distortion (RSD) correction. 
We estimate the polarization signal with realistic modeling of the ionization power spectrum using the seminumerical simulation results. 
The formalism is kept in terms of the angular power spectrum, $C_\ell$, instead of the power spectrum, $P(\vec k)$,
since the former approach is more adapted to multifrequency studies and wide-field surveys 
\citep{2004ApJ...608..622Z, 2007MNRAS.378..119D, 2016ApJ...833..242L}. 
We will investigate new features in the polarization angular power spectra, which may be used for constraining reionization.
On the other hand, we will also reevaluate the detection prospects of the polarization signal by the upcoming SKA telescope, including a brief discussion on the effects of foregrounds, systematics, and Faraday rotation.

The rest of this paper is organized as follows. 
In \S\ref{sec:deltaTb}, we reformulate the angular power spectra of 
the 21~cm linear polarization signal due to Thomson scattering, based on a fully relativistic framework, 
while leaving the detailed derivations to 
Appendices \ref{app:deltaTb}, \ref{app:generalCl}, and \ref{app:signalcompare}.
We then describe the EoR modeling and our seminumerical simulations in \S\ref{sec:model} 
which provide the initial source fields for the 21~cm signal.
In \S\ref{sec:results}, we present the numerical results and discuss their cosmological implications.
We briefly discuss the observational prospects of the polarization signal in \S\ref{sec:detection}, 
and make concluding remarks in \S\ref{sec:conclusions}.

\section{The 21~cm linear polarization} \label{sec:deltaTb}

\subsection{Temperature anisotropy} \label{ssec:deltaTb}

Before showing the 21~cm temperature anisotropy, 
we briefly clarify all the approximations made in this paper.
\begin{enumerate} 
\item[(1)] The 21~cm lines are optically thin, $\tau_{\nu_{\rm obs}}\ll 1$ \citep[e.g.,][]{2007PhRvD..76h3005L}.
\item[(2)] The IGM has been preheated so that
$T_{\rm s}\gg T_{\rm CMB}$.
It is valid soon after reionization begins, when $T_{\rm s}$ quickly reaches above $10^4$ K 
because of sufficient heating of the IGM due to X-ray photons, and efficient coupling of $T_{\rm s}$ with the kinetic temperature of the gas via Ly$\alpha$ pumping
\citep[e.g.,][]{2004ApJ...602....1C,2006MNRAS.371..867F, 2009A&A...495..389B}. 
\item[(3)] During the EoR, matter fluctuations are still Gaussian and linear on most relevant scales, $|\delta|\ll1$, 
whereas H~{\small I} density fluctuations are not, due to reionization patchiness. 
This is the so-called ``quasi-linear'' regime \citep{2012MNRAS.422..926M}.
\end{enumerate}
Altogether, it is reasonable to assume the 
\emph{optically-thin, post-heating, quasi-linear} regime
for most emitting H~{\small I} gases during the EoR\footnote{The 21~cm signal from a fixed redshift or $\nu_{\rm obs}$ is also subject to the light-cone (LC) effect \citep[e.g.,][]{2006MNRAS.372L..43B}. To exactly account for the LC anisotropy, one must resort to full numerical schemes 
\citep[e.g.,][]{2012MNRAS.422..926M, 2018MNRAS.474.1390M, 2019MNRAS.490.1255C}. However, the LC effect is important only for the longitudinal modes along the LoS, but our focus is the transverse 2D modes perpendicular to the LoS. Therefore, in this paper we will neglect the LC effect for the EoR signal, evaluating it at the ensemble-averaged radial distance of 21~cm emission events
from all directions. 
Justification of this treatment can be found in more detail in Appendix \ref{app:deltaTbz}.}. 
We refer interested readers to the detailed implications of these assumptions in Appendix \ref{app:assumption}.

Under these approximations, we can model the 21~cm brightness temperature 
at the observed frequency $\nu_{\rm obs}$ 
along the direction of $\hat n$ as seen by an observer
at $(\eta_{\rm obs}, \vec x)$, where $\eta$ is the conformal time, as 
\begin{equation}\label{eq:tbpostheating}
\delta T_{\rm b} \approx  T_0(z) \bar x_{\rm HI, m}(z_{\rm em})
	\Big(1+\delta_{\rm HI}\Big)\left(1-\frac{1}{\mathscr{H}}\frac{\partial v_\parallel}{\partial r}\right)\,.
\end{equation}
In this expression, the terms responsible for the global signal include 
the \emph{mass-weighted} average neutral fraction, 
$\bar x_{\rm HI, m}\equiv\bar n_{\rm HI}/\bar n_{\rm H}$, at the redshift of emission $z_{\rm em}$, 
and the dimensional factor $T_0(z)$, 
which we refer readers to Appendix \ref{app:deltaTbdetail} for its definition.
The fluctuations of the signal are determined by 
$\delta_{\rm HI}$ (the neutral hydrogen density fluctuations), and the LoS velocity gradient term, 
$(1/\mathscr{H})(\partial v_\parallel / \partial r)$, which accounts for the RSD effect, where $v_\parallel\equiv\vec v\cdot\hat r=-\vec v\cdot \hat n$ 
is the LoS projection of the H~{\small I} peculiar velocity at emission, 
and $\mathscr{H}\equiv aH$ is the conformal Hubble parameter.
The RSD term is small in the quasi-linear regime, 
$\left|(1/\mathscr{H})(\partial v_\parallel / \partial r)\right|\sim|\delta| \ll 1$. Here $\delta$ is the gauge-dependent matter overdensity.

Our primary interest is the \emph{fluctuations and anisotropies} of the signal. 
We define the dimensionless brightness temperature as 
$\psi (\eta_{\rm obs}, \vec x, \nu_{\rm obs}, \hat n) \equiv \delta T_{\rm b}(\eta_{\rm obs}, \vec x, \nu_{\rm obs}, \hat n) / T_0(z)$. 
The fluctuations and anisotropies of $\psi$ are defined as
$\Theta(\eta_{\rm obs}, \vec x,\nu_{\rm obs}, \hat n)\equiv\psi/\langle\psi\rangle - 1$. In Fourier space, its conjugate is (see the derivation in Appendix \ref{app:deltaTbdetail})
\begin{equation} \label{eq:tbkfinal}
	\Theta (\eta_{\rm obs}, \vec k, \nu_{\rm obs}, \mu) \approx \delta_{\rm HI}(\vec k, z_{\rm em})e^{-i\mu ks} + \mu^2\delta(\vec k, z_{\rm em})e^{-i\mu ks},
\end{equation} 
where $\mu\equiv\hat k\cdot \hat n$ and $s$ is the comoving radial distance in the redshift space. 

The free-streaming projection of plane waves is manifest in Eq.~(\ref{eq:tbkfinal}).
Consequently, multipole moments of temperature anisotropies 
can be defined in the $\hat z=\hat k$ frame (so that $\mu=\cos\theta$):
\begin{IEEEeqnarray}{rl}\label{eq:Thetaell}
	\Theta_\ell(\eta_{\rm obs}, \vec k, \nu_{\rm obs}) \equiv & 
	\frac{1}{(-i)^\ell}\int^1_{-1}\frac{\ud\mu}{2}\mathscr{P}_\ell(\mu)\Theta(\eta_{\rm obs}, \vec k, \nu_{\rm obs}, \mu)\nonumber\\
	= \delta_{\rm HI}(\vec k, z_{\rm em}) & j_\ell(ks)-\delta(\vec k, z_{\rm em})j_\ell^{''}(ks),
\end{IEEEeqnarray}
where $\mathscr{P}_\ell(\mu)$ is the Legendre polynomial, 
$j_\ell(x)$ is the spherical Bessel function, and $j_\ell''(x)$ is the second-order derivative of $j_\ell(x)$ with respect to its argument. 

\subsection{Linear polarization from electron scattering} \label{ssec:polarization}

Thomson scattering couples the intensity to the $E$-mode polarization,
thereby generating the latter out of the unpolarized light.
For scalar perturbations, the $B$-mode polarization is decoupled from the $E$-mode 
and vanishes under the azimuthal symmetry around $\hat k$. 
Analogous to the CMB, the transport of the 21~cm polarization signal 
follows the Boltzmann equations 
and its observed anisotropies can be calculated by the standard LoS integration formalism
(\citealt{1997PhRvD..55.1830Z}; BL05). 
It exploits the multipole expansion (Eq.~\ref{eq:Thetaell}) in the $\hat z = \hat k$ frame 
and yields the present-day observed (scalar) EE angular power spectrum, which is defined as 
$C^{\rm EE}_\ell (\eta_0, \nu_{\rm obs,0})   \equiv  \langle |a_{{\rm E}, \ell m}(\eta_0, \vec x_0, \nu_{\rm obs,0})|^2\rangle $, 
where $a_{{\rm E}, \ell m}$ are coefficients of the spherical harmonics decomposition. 
For a present-day Earth observer, 
\begin{equation}
	C^{\rm EE}_\ell (\eta_0, \nu_{\rm obs,0})   
	=  4\pi\bigg\langle\left| \int\frac{\ud^3\vec k}{(2\pi)^3}\Delta_{{\rm E}\ell}(\eta_0, \vec k, \nu_{\rm obs,0})\right|^2\bigg\rangle.
\end{equation}
In the expression above, $\Delta_{{\rm E}\ell}$ are multipole moments of $E$-mode polarization anisotropies in the $\hat z=\hat k$ frame, 
and are derived in the standard LoS integration formalism. 
Here we write down its expression as follows:
\begin{IEEEeqnarray}{rl}	\label{eq:E_ell}
	& \Delta_{{\rm E}\ell} (\eta_0, \vec k, \nu_{\rm obs,0}) \approx \frac{3}{4}\sqrt{\frac{(\ell+2)!}{(\ell-2)!}} \nonumber\\
	& \times\int^{\bar\eta_0}_{\bar\eta_{z_{\rm em}}} g(\eta')\Pi(\eta', \vec k, \nu')\frac{j_\ell[ck(\bar\eta_0-\eta')]}{[ck(\bar\eta_0-\eta')]^2} \ud\eta',\quad
\end{IEEEeqnarray}
where $\nu'=\nu_{\rm obs,0}(1+z')\leq \nu_{21}$ is the frequency of 21~cm photons
seen by an intervening free electron, 
and the source function $\Pi(\eta', \vec k, \nu')
=\Theta_2(\eta', \vec k, \nu')+\Delta_{P2}(\eta', \vec k, \nu')+\Delta_{P0}(\eta', \vec k, \nu')$,
where $\Theta_2$ is the quadrupole moment of temperature fluctuations, 
and $\Delta_{{\rm P}\ell}$ are multipole moments of the total linear polarization.
The (global) visibility function is defined as $g(\eta)\equiv-(\ud\tau/\ud\eta) e^{-\tau}$, 
where $\tau$ is the Thomson scattering optical depth, 
$\tau(\eta)\equiv\int^{\eta_0}_{\eta}\bar n_{\rm e}(\eta')\sigma_{\rm T}c\,\ud\eta'$, 
$\sigma_{\rm T}$ is the Thomson scattering cross section.
$g(\eta)$ depends on the global ionization history, illustrated in \S\ref{ssec:simulation}.
In Eq.~(\ref{eq:E_ell}), the density of free electrons
is approximated by the globally-averaged value, $\bar n_{\rm e}$.
However, in reality their distribution is patchy during the EoR
and may cause secondary polarization 
\citep[see, e.g.,][for the case of the CMB]{2000ApJ...529...12H, 2007PhRvD..76d3002D}.
This effect is worth further examinations 
but for now we neglect it as in BL05.

Given the low value of the electron scattering optical depth ($\tau_{\rm es}\sim0.05$), 
it is reasonable to assume that most of the 21~cm photons do not scatter more than once 
by free electrons before they reach the observer.
As a result, the relevant source function $\Pi$ for 21~cm polarization (with the CMB part subtracted off) 
only has the contribution from the incident quadrupole, $\Theta_2$. 
(Note that this is not the case for the CMB in the tight coupling limit).
Also, under the current sensitivity of 21~cm experiments, 
the nuisance that the 21~cm temperature signal 
is suppressed by a factor of $\exp{(-\tau_{\rm es})}$ due to Thomson scattering can be neglected
so that $\Theta(\vec k,\mu)$ remains at the free-streaming value (Eq.~\ref{eq:tbkfinal}).

On the other hand, $\Theta$ from Eq.~(\ref{eq:tbkfinal}) can be viewed as a decomposition 
according to the initial source fields, $\delta_{\rm HI}$ and $\delta$, 
which are statistically homogeneous and isotropic.
Each term has its own temperature transfer function: 
$T^{\rm T}_{\rm HI}(\eta_{\rm obs}, k, \nu_{\rm obs}, \mu)=e^{-i\mu k s}$
and $T^{\rm T}_\delta(\eta_{\rm obs}, k, \nu_{\rm obs}, \mu)=\mu^2 e^{-i\mu k s}$.
Evidently, Eq.~(\ref{eq:Thetaell}) implies that 
multipole moments of these temperature transfer functions are
$T^{\rm T}_{{\rm HI},\ell}(\eta_{\rm obs}, k, \nu_{\rm obs})=j_l[ck(\bar\eta_{z_{\rm obs}}-\bar\eta_{z_{\rm em}})]$
and $T^{\rm T}_{\delta,\ell}(\eta_{\rm obs}, k, \nu_{\rm obs})=-j_l''[ck(\bar\eta_{z_{\rm obs}}-\bar\eta_{z_{\rm em}})]$.
\begin{widetext}
Thus, the present-day EE power spectrum can be explicitly written as
\begin{equation}\label{eq:EEps}
	C^{\rm EE}_\ell (\eta_0, \nu_{\rm obs,0})  = \frac{2}{\pi}\int k^2\ud k
	\Big[P_{\rm HI}(k, z_{\rm em})\left(\mathscr{T}^{\rm E}_{{\rm HI},\ell}(k) \right)^2 
	+ P_\delta(k, z_{\rm em})\left(\mathscr{T}^{\rm E}_{\delta,\ell}(k) \right)^2 
	+ 2P_{{\rm HI}\,\delta}(k, z_{\rm em})\mathscr{T}^{\rm E}_{{\rm HI},\ell}(k)
	\mathscr{T}^{\rm E}_{\delta,\ell}(k) \Big].
\end{equation}
Here the multipole moments of the $E$-mode polarization transfer functions in the integrand of the RHS of Eq.~(\ref{eq:EEps}), as well as those temperature transfer functions in the integrand of the RHS of Eq.~(\ref{eq:TTps}--\ref{eq:TEps}) below, are all implicitly evaluated at $(\eta_0, \nu_{\rm obs,0})$, 
\begin{subequations}\label{eq:EtransLoS}
\begin{IEEEeqnarray}{rCl}
	\mathscr{T}^{\rm E}_{{\rm HI},\ell}(\eta_0, k, \nu_{\rm obs,0}) & \equiv & \frac{3}{4}\sqrt{\frac{(\ell+2)!}{(\ell-2)!}}
	\int^{\bar\eta_0}_{\bar\eta_{z_{\rm em}}}g(\eta')j_2[ck(\eta'-\bar\eta_{z_{\rm em}})]
	\frac{j_\ell[ck(\bar\eta_0-\eta')]}{[ck(\bar\eta_0-\eta')]^2}\ud \eta', \\
	\mathscr{T}^{\rm E}_{\delta,\ell}(\eta_0, k, \nu_{\rm obs,0}) & \equiv & -\frac{3}{4}\sqrt{\frac{(\ell+2)!}{(\ell-2)!}}
	\int^{\bar\eta_0}_{\bar\eta_{z_{\rm em}}}g(\eta')j''_2[ck(\eta'-\bar\eta_{z_{\rm em}})]
	\frac{j_\ell[ck(\bar\eta_0-\eta')]}{[ck(\bar\eta_0-\eta')]^2}\ud \eta'.
\end{IEEEeqnarray}
\end{subequations}
Here $P_{\rm HI}$ and $P_\delta$ are the (equal-time) H~{\small I} density and matter power spectra, respectively, 
and $P_{{\rm HI}\,\delta}$ is the cross-power spectrum between the H~{\small I} and the matter field.
They are defined as $\langle \delta^*_{\rm HI}(\vec k)\delta_{\rm HI}(\vec k')\rangle\equiv(2\pi)^3P_{\rm HI}(k)\delta^{(3)}_{\rm D}(\vec k-\vec k')$,
$\langle \delta^*(\vec k)\delta(\vec k')\rangle\equiv(2\pi)^3P_\delta(k)\delta^{(3)}_{\rm D}(\vec k-\vec k')$,
and $\langle \delta^*_{\rm HI}(\vec k)\delta(\vec k')\rangle\equiv(2\pi)^3P_{{\rm HI}\,\delta}(k)\delta^{(3)}_{\rm D}(\vec k-\vec k')$.
Finally, to obtain dimensional quantities, 
the extra coefficient of $\left[T_0(z_{\rm em})\bar x_{\rm HI,m}(z_{\rm em})\right]^2$ 
should be multiplied to the result of the dimensionless angular power spectrum.

Similarly, the temperature power spectrum observed today, 
$C^{\rm TT}_\ell = 4\pi\Big\langle\left| \int\ud^3\vec k \,\Theta_\ell(\eta_0, k, \nu_{\rm obs,0})/(2\pi)^3\right|^2\Big\rangle$, is given by
\begin{equation} \label{eq:TTps}
	C^{\rm TT}_\ell (\eta_0, \nu_{\rm obs,0})
	= \frac{2}{\pi}\int k^2\ud k\Big[P_{\rm HI}(k, z_{\rm em})\left(T^{\rm T}_{{\rm HI},\ell}(k) \right)^2 
	+ P_\delta(k, z_{\rm em})\left(T^{\rm T}_{\delta,\ell}(k) \right)^2 
	+ 2P_{{\rm HI}\,\delta}(k,z_{\rm em})T^{\rm T}_{{\rm HI},\ell}(k) T^{\rm T}_{\delta,\ell}(k) \Big].
\end{equation}
The cross-power spectrum between the temperature and the $E$-mode polarization is 
\begin{IEEEeqnarray}{rl}  \label{eq:TEps}
	C^{\rm TE}_\ell (\eta_0, \nu_{\rm obs,0}) =  \frac{2}{\pi}\int k^2\ud k
	\Big\{ & P_{\rm HI}(k, z_{\rm em})T^{\rm T}_{{\rm HI},\ell}(k) \mathscr{T}^{\rm E}_{{\rm HI},\ell}(k)
	+ P_\delta(k, z_{\rm em})T^{\rm T}_{\delta,\ell}(k) \mathscr{T}^{\rm E}_{\delta,\ell}(k) \nonumber\\
     	& + P_{{\rm HI}\,\delta}(k, z_{\rm em})\big[T^{\rm T}_{{\rm HI},\ell}(k)\mathscr{T}^{\rm E}_{\delta,\ell}(k)
	+T^{\rm T}_{\delta,\ell}(k)\mathscr{T}^{\rm E}_{{\rm HI},\ell}(k)\big]\Big\}.
\end{IEEEeqnarray}
\end{widetext}

Eqs.~(\ref{eq:EEps}) and (\ref{eq:TEps}) describe the polarization signals that we seek in this paper. 
The case in which 21~cm brightness temperature fluctuations 
are sourced by multiple cosmological fields in a more general context 
is formulated in Appendix~\ref{app:generalCl}.

\subsection{Previous modeling of the 21~cm polarization} \label{ssec:comparison}

An equivalent way to write down the dimensionless brightness temperature is 
\begin{IEEEeqnarray}{rCl}\label{eq:ndTb}
	\psi  & \equiv & x_{\rm HI}\left[1+\Delta \right] =  x_{\rm HI}+\bar x_{\rm HI}\left[\Delta+\delta x_{\rm HI}\Delta \right],
\end{IEEEeqnarray}
where $\Delta\equiv\delta-(1/\mathscr{H})(\partial v_\parallel/\partial r)$ and $x_{\rm HI}$ is the local neutral fraction. 
It is implicitly assumed that the baryon distribution follows that of the total matter.
BL05 ignored the nonlinear coupling term $\delta x_{\rm HI}\Delta$, so that they approximated $\psi \approx x_{\rm HI}+\bar x_{\rm HI}\Delta$ 
and calculated the temperature multipole $\bar x_{\rm HI, m}(z_{\rm em}) \Theta_\ell(\eta_{\rm obs}, \vec k, \nu_{\rm obs})$ as 
\begin{equation}\label{eq:quadTb}
	\bar x_{\rm HI}(z_{\rm em}) \Big[ \delta x_{\rm HI}(\bar\eta_{z_{\rm em}}, \vec k)j_\ell(ks) 
	+ \delta(\bar\eta_{z_{\rm em}}, \vec k)\Big(j_\ell(ks)-j_\ell^{''}(ks)\Big)\Big].\qquad
\end{equation} 
Also, the cross-power between the two terms on the RHS of 
Eq.~(\ref{eq:quadTb}) was missing from their modeling.

Nevertheless, both nonlinear term $\delta x_{\rm HI}\Delta$ and the cross-power $P_{x_{\rm HI}\,\delta}$ are non-negligible, since fluctuations in the ionization field is significant ($\delta x_{\rm HI}\sim 1$), as will be shown in \S\ref{ssec:HIbias} (see Figure~\ref{fig:Pk}). 
Our approach improves upon previous modeling by including not only the autocorrelations of the source fields (the ionization field and the corrections for the RSD), but also the contributions from cross-correlations $P_{{\rm HI}\,\delta}(k)$. 
We leave a detailed comparison of formulations to Appendix~\ref{app:signalcompare}.

\section{Reionization Modeling} \label{sec:model}

In the previous section, we focused on the accurate modeling 
of the 21~cm linear polarization signal from the EoR.
In this section, we focus on modeling the inhomogeneous reionization that sources the 21~cm signal.

Generally, cosmic reionization began when the first structures formed around $z\approx 30$
\citep[e.g.,][]{2001PhR...349..125B, 2018PhR...780....1D}. 
Each luminous source first creates an ionized region (``H~{\small II} bubbles'') around itself.
Ionized regions grow and later overlap, percolating into the IGM. 
The completion of bubble overlapping marks the end of reionization
by $z\sim 5-6$ \citep[e.g.,][]{2010MNRAS.407.1328M}.
For 21~cm polarization, the dominant source for the EoR signal is the ionization field, 
or equivalently, the neutral fraction field,
$\delta x_{\rm HI}\equiv x_{\rm HI}/\bar x_{\rm HI}-1$
($\bar x_{\rm HI}$ is the \emph{volume-weighted} average neutral fraction).
It is patchy and nonlinear during the bulk of cosmic reionization, 
as supported by observations \citep{2015MNRAS.447.3402B, 2017MNRAS.466.4239G}.

There are a variety of approaches to model the inhomogeneous ionization field,
including analytical, seminumerical and fully-numerical methods.
Early analytical models often approximate two-point statistics of the ionization field with simplified, yet physically motivated ansatz
\citep[e.g,][]{1998ApJ...508..435G, 1998PhRvL..81.2004K}.
For example, BL05 employed a simple bubble model to estimate the neutral fraction power spectrum,
which assumes that H~{\small II} regions are randomly-distributed, fully-ionized spheres. 
\citet{2004ApJ...613....1F} introduced more sophisticated analytical model that is based on the excursion set formalism.  
This formalism relates the emissivity of ionizing sources to the underlying matter distribution.
Based upon this scheme, efficient \emph{seminumerical} algorithms 
have been developed to generate the realizations of ionization fields 
without the radiative transfer computations 
\citep[e.g.,][]{2007ApJ...669..663M, 2010MNRAS.406.2421S,2011MNRAS.411..955M}. 
Finally, to provide the most accurate modeling of reionization, 
one needs cosmological radiative transfer simulations and/or radiation-hydrodynamic simulations 
with large enough dynamical ranges and reasonable prescriptions for ionization sources
\citep[e.g.,][]{2006MNRAS.369.1625I, 2007MNRAS.377.1043M, 
2007ApJ...671....1T,2010A&A...523A...4B, 2013ApJ...776...81B, 2013MNRAS.436.2188R, 
2014MNRAS.439..725I, 2015ApJ...807L..12O, 2016MNRAS.463.1462O, 2020MNRAS.491.1600M}.

In this paper, we employ the seminumerical approach to estimate the ionization power spectra. 
It can model the effect of patchy reionization more realistically than analytical methods, 
and is computationally more efficient than fully-numerical methods. 
Moreover, comparisons between radiative transfer simulations 
and seminumerical simulations of reionization 
\citep[e.g.,][]{2007ApJ...654...12Z, 2011MNRAS.414..727Z, 2014MNRAS.443.2843M, 2018MNRAS.477.1549H} 
demonstrated that their predictions on the power spectrum agree with each other in reasonable accuracies at the scales of interest to upcoming interferometric observations. 
Thus, their difference should not affect the conclusions presented in this paper.

We present our seminumerical simulations in \S\ref{ssec:simulation}, briefly review early analytical models in \S\ref{ssec:toymodel}, and discuss the H~{\small I} power spectrum and H~{\small I} bias in \S\ref{ssec:HIbias}.

\subsection{Seminumerical reionization simulations} \label{ssec:simulation} 

The density fields and ionization fields are generated
from cosmological seminumerical simulations of reionization using the {\tt 21cmFAST} code\footnote{\url{https://github.com/andreimesinger/21cmFAST}} \citep{2011MNRAS.411..955M},
with two choices of comoving boxes: 
the large box with 2000 cMpc per side 
and the small box with 512 cMpc per side.
For the large (small) box, initial conditions are created 
on a $4000^3$ ($4096^3$) grid, 
and smoothed down to a $1000^3$ ($1024^3$) grid
for the reionization simulation.
We output data at desired redshifts during the post-heating phase of the EoR,
ranging from $z=10$ to $z=5.6$.

The simulations are based on three fundamental parameters in the EoR modeling: $\zeta$ (ionization efficiency), $T_{\rm vir}$ (the minimum virial temperature of halos that host ionizing sources), and $R_{\rm mfp}$ (the maximum mean free path of ionizing photons).
Detailed astrophysical interpretations of these parameters and 
how the global ionization history depends on them
can be found in \citet{2011MNRAS.411..955M,2017MNRAS.465.4838G,2018MNRAS.477.3217G,2020JOSS....5.2582M}.

For both the large and small boxes, 
we apply multiple sets of EoR parameters, 
labeled as ``Model 1-3'' in Table \ref{tab:EoRparam}.\footnote{In this paper, we assume a $\Lambda$CDM cosmology with $\Omega_{\rm m} = 0.32$, $\Omega_{\rm b}=0.05$, $h=0.67$, $n_{\rm s}=0.97$ and $\sigma_8=0.81$, in consistent with the \citetalias{2020A&A...641A...6P}.}
They generate different global ionization histories, 
and thus different electron scattering optical depths 
and visibility functions (defined in \S\ref{ssec:polarization}).
For each model in Table~\ref{tab:EoRparam}, 
we have verified that the global ionization histories from the two box sizes are consistent.

\begin{figure}
\resizebox{\columnwidth}{!}{\includegraphics{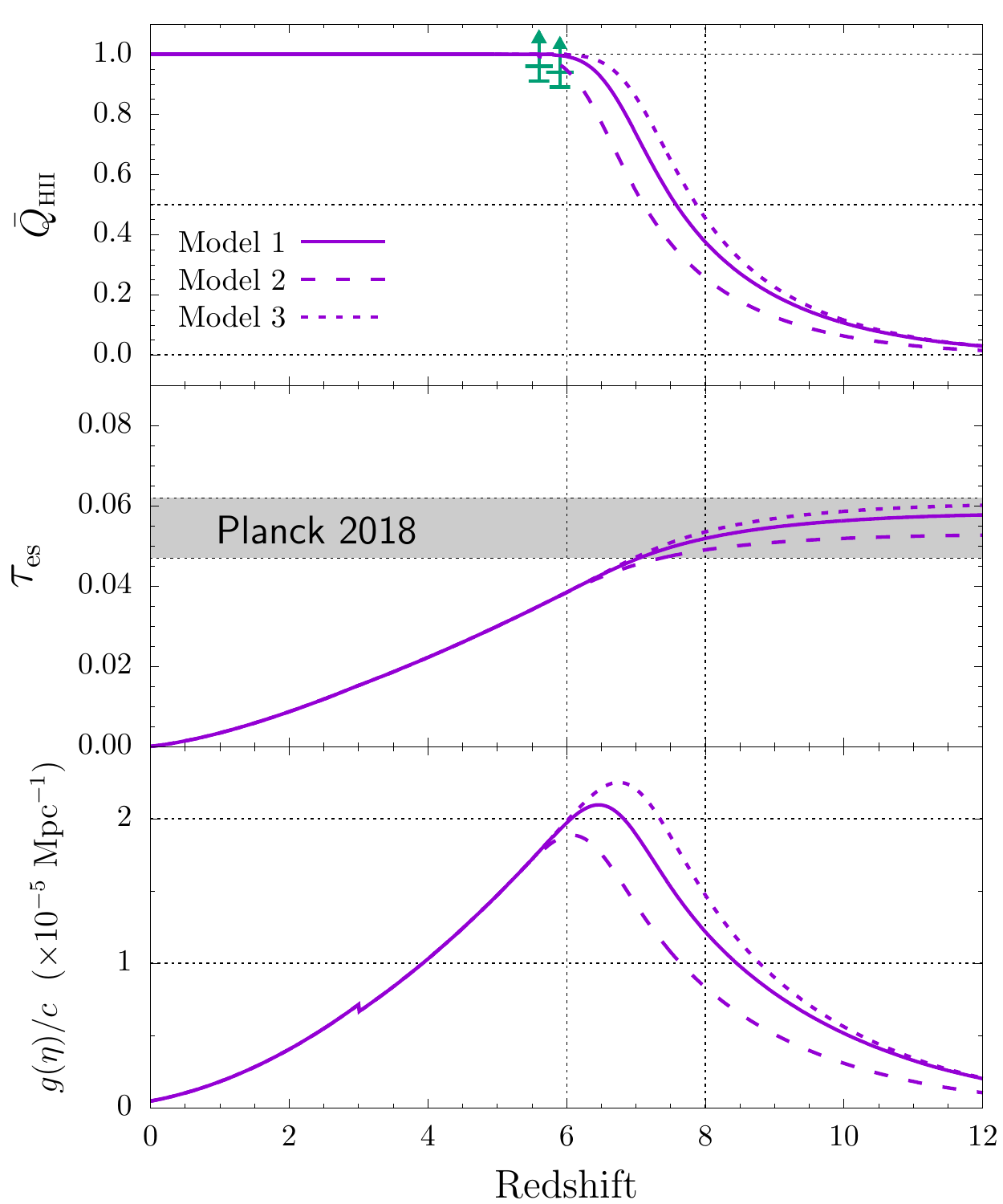}}
\caption{Reionization histories generated by the seminumerical simulations of reionization with EoR model parameters $(\zeta,T_{\rm vir},R_{\rm mfp})$ in Table \ref{tab:EoRparam} (solid/dashed/dotted curves for Model 1/2/3). Shown are the H~{\small II} volume filling fraction $\bar Q_{\rm H {\tiny II}}$ (top), the CMB Thomson scattering optical depth $\tau_{\rm es}$ (middle), and the visibility function normalized by $c$, $g(\eta)/c$ (bottom), as a function of redshift. These reionization histories are all consistent with 1$\sigma$ constraints from the fraction of ``dark'' pixels in the Ly$\alpha$ and Ly$\beta$ forests
\citep[][arrows in the top panel]{2015MNRAS.447..499M}
and the global $\tau_{\rm es}$ inferred by the CMB
\citepalias[][the shaded area in the middle panel]{2020A&A...641A...6P}.  
In the bottom panel, the discontinuity at $z=3$ is due to He~{\scriptsize II} reionization.}
\label{fig:EoRhistory}
\end{figure}

Figure~\ref{fig:EoRhistory} shows the ionization histories,
compared with current model-independent observational constraints from
(1) the fraction of ``dark'' pixels in the Ly$\alpha$ and Ly$\beta$ forests,
and (2) the global Thomson optical depth inferred by the CMB.
$\bar Q_{\rm H {\tiny II}}\equiv 1-\bar x_{\rm HI}$
defines the H~{\small II} volume filling fraction (a.k.a.~the global ionized fraction).
The top and middle panels verify the physical dependence of the ionization history 
on the EoR parameters. For example, Model~2 has the highest $T_{\rm vir}$ among the three, 
resulting in the latest beginning of reionization and the lowest value of $\tau_{\rm es}$. 
For Model~3, it has a higher ionization efficiency $\zeta$ (and a slightly higher $T_{\rm vir}$) than Model~1, 
which leads to the earliest onset and end of reionization among the three.
The bottom panel of Figure~\ref{fig:EoRhistory} shows that the visibility function,
while reaching its maximum near the end of the EoR, 
demonstrates a broad width over redshift, 
in contrast to the sharply-peaked $g(\eta)$ at recombination with respect to the CMB
\citep{1995ApJ...444..489H}.
Therefore, the LoS integration in Eqs.~(\ref{eq:EtransLoS}a--\ref{eq:EtransLoS}b) 
is more time-consuming than in the CMB case.

\begin{deluxetable}{cccc}
\tablenum{1}
\tablecaption{EoR model parameters\label{tab:EoRparam}}
\tablewidth{0pt}
\tablehead{
\colhead{Model} & \colhead{$\zeta$} & \colhead{$T_{\rm vir}$ [K]} & \colhead{$R_{\rm mfp}$ [cMpc]} 
}
\startdata
1 & 30 & $6\times10^4$ & 40 \\
2 & 35 & $1\times10^5$ & 40 \\
3 & 40 & $7\times10^4$ & 35 \\
\enddata
\end{deluxetable}

\subsection{Analytical models} \label{ssec:toymodel} 

For the purpose of comparison, here we review some of the early analytical models for the ionization field.
BL05 made an assumption that at the end of the EoR (at $z_{\rm re}$), 
H~{\small II} regions are approximately Poisson-randomly-distributed bubbles (BL05; \citealt{2014PhRvD..89l3002D}).
As a result, fluctuations in the ionized fraction are dominated by those in the local number of bubbles. 
Also, reionization is assumed to be instantaneous at $z_{\rm re}$.
Based on these assumptions, the ionization power spectrum $P_x(k)$ 
(i.e.\ power spectrum of the neutral fraction field $x_{\rm HI}$) 
is estimated with the following ansatz at the end of the EoR ($z_{\rm em} =z_{\rm re}$),
\begin{equation}\label{eq:toymodelBL05}
	P_x(k)=\frac{4\pi R^3}{3}e^{-k^2R^2},
\end{equation}
where $R$ is the characteristic radius of bubbles at $z_{\rm re}$, 
$R\propto (1+z_{\rm re})^{-3/2}$ \citep{2004Natur.427..815W}.
However, this power spectrum is not realistic 
since $P_x(k)$ should vanish when reionization ends $(\bar x_{\rm HI}\to 0)$.
Therefore, in this paper, we will not compare our results with toy model in Eq.~(\ref{eq:toymodelBL05}).

A more realistic analytical model was given by \citet{2004ApJ...608..622Z} 
which also assumes randomly-distributed, single-size bubbles, 
but is not restricted to the end of the EoR,
satisfying the constraint that $P_x(k)\to 0$ when reionization finishes. 
In their model, 
\begin{equation} \label{eq:ZFH04model}
	P_x(k)=(\sqrt{2\pi}R)^3(\bar x_{\rm HI}-\bar x_{\rm HI}^2)e^{-k^2R^2/2}.
\end{equation}
In this paper, we will compare our simulation results with this analytical model (Eq.~\ref{eq:ZFH04model}).

\begin{figure}
\resizebox{\columnwidth}{!}{\includegraphics{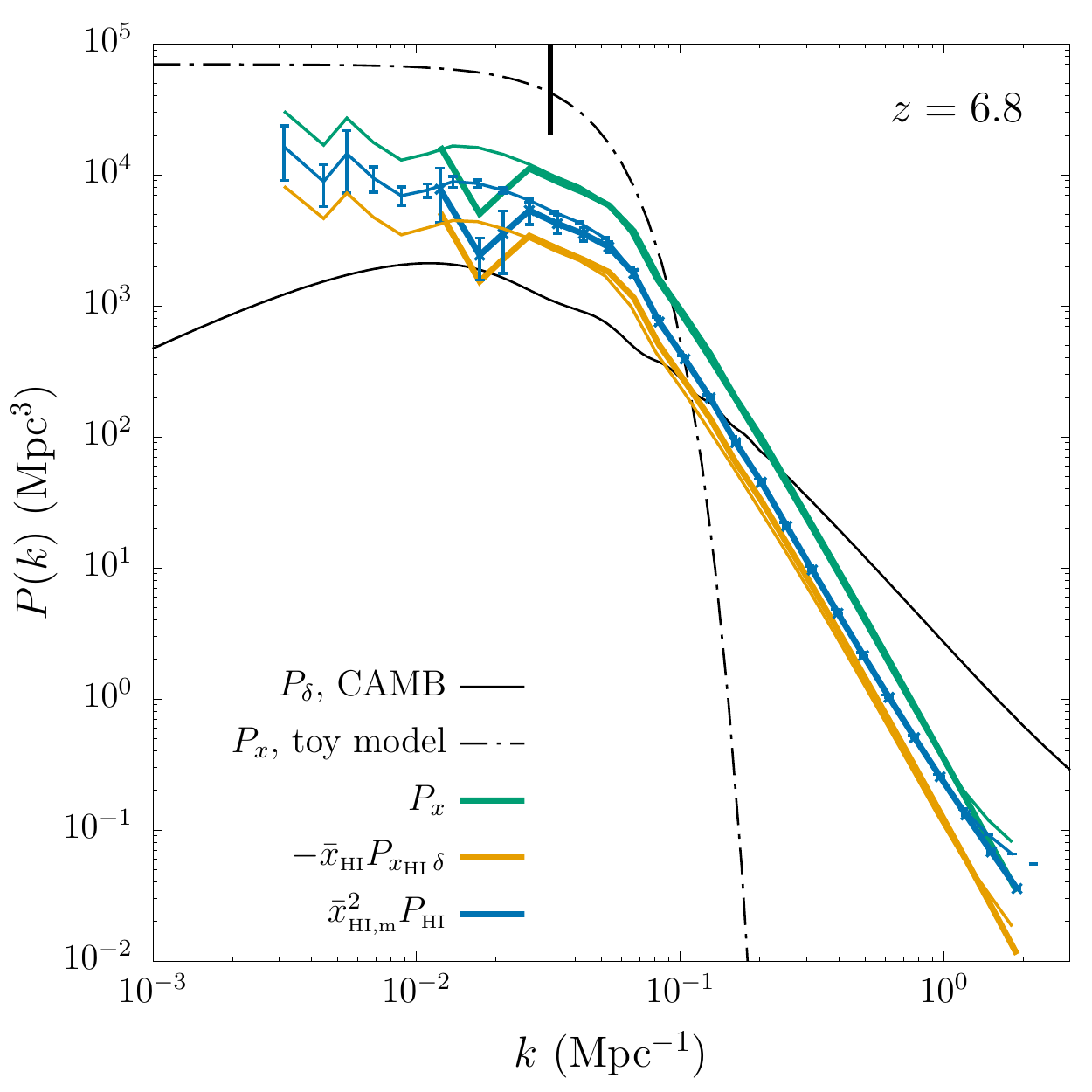}}
\caption{Power spectra of various sources at $z=6.8$ (corresponding to $\bar Q_{\rm H {\tiny II}}\approx 0.82$ in Model 1). 
Shown are the neutral fraction power spectrum $P_x(k)$ (green solid line), the H~{\scriptsize I} density power spectrum $P_{\rm HI}(k)$ (blue solid line), and the neutral fraction-density cross-power spectrum $P_{x_{\rm HI}\,\delta}(k)$ (yellow solid line), with the latter two power spectra being weighted by relevant factors. These power spectra are calculated from the seminumerical simulations of reionization in the small box ($512$ cMpc per side, thick colored lines), and in the large box ($2000$ cMpc per side, thin colored lines). 
The error bars on $P_{\rm HI}$ represent the sample variances corresponding to each simulation volume. 
For the purpose of comparison, we also show the matter power spectrum produced by {\tt CAMB} (black solid line), 
and the neutral fraction power spectrum $P_x(k)$ from the toy model using Eq.~(\ref{eq:ZFH04model}) (dot-dashed line) with the characteristic bubble radius in this toy model indicated by the black thick vertical line.  
}
\label{fig:Pk}
\end{figure}

\begin{figure*}
\resizebox{\textwidth}{!}{\includegraphics{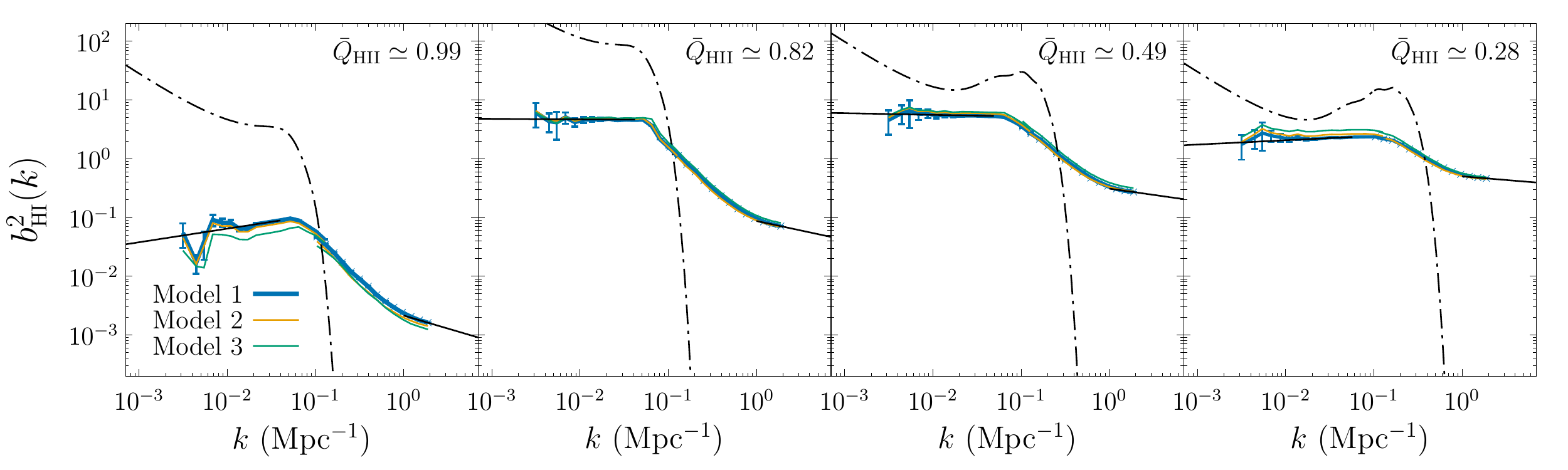}}
\caption{The H~{\scriptsize I} bias at different ionized fractions during the EoR. Shown are the results from seminumerical reionization simulations for all three models in Table~\ref{tab:EoRparam} (blue error bars representing the sample variance from Model 1), and those from the toy model (Eq.~\ref{eq:ZFH04model}, dot-dashed lines). We extrapolate the bias from simulation with power-law fits (black solid lines) at both high and low ends of the $k$ range beyond the simulation coverage.
}
\label{fig:HIbias}
\end{figure*}

\subsection{H~{\small I} power spectrum and bias} \label{ssec:HIbias}

In Figure~\ref{fig:Pk}, we illustrate the (equal-time) power spectrum of the H~{\small I} density field, $P_{\rm HI}(k)$,
and that of the neutral fraction field, $P_x(k)$, both obtained from simulation data.
The matter power spectrum is calculated using the {\tt CAMB} package\footnote{\url{https://camb.info/}}.
The simulated power spectra -- $P_{\rm HI}(k)$, $P_x(k)$, and the cross-power $P_{x_{\rm HI}\,\delta}(k)$ 
-- have similar shapes on all scales of interest, 
which indicates that the neutral fraction fluctuations dominate over the density fluctuations. 
For the purpose of a fair comparison between the simple analytical model 
and the seminumerical simulations, we set the characteristic bubble radius at the end of reionization 
in the toy model (Eq.~\ref{eq:ZFH04model}) to be the photon horizon in the simulation, 
and take into account its growth during the EoR. 
On scales smaller than the characteristic bubble radius (sub-bubble scales), 
realistic $P_x(k)$ from simulation shows a power-law-like decrease,
slower than the exponential cutoff from the toy model.
On scales larger than the characteristic bubble radius (super-bubble scales), however, 
the simulated power spectrum is smaller than the estimate from the toy model by nearly one order of magnitude.
The reason is that at any fixed global ionized fraction during the EoR,
realistic H~{\small II} regions can have internal structures, 
whereas the toy model assumes fully ionized bubbles (two-phased IGM). 
Such a simplification would effectively transfer powers from sub-bubble scales to super-bubble scales.

Moreover, Figure~\ref{fig:Pk} clearly shows that 
the neutral fraction and the matter overdensity field are anti-correlated, 
as a result of the inside-out reionization scenario.
Also, their cross-power spectrum, $P_{x_{\rm HI}\,\delta}(k)$, 
has comparable amplitude to $P_x(k)$, and hence
is a non-negligible contribution to the 21~cm polarization power spectrum. 
Therefore, we include the cross-power ($P_{{\rm HI}\,\delta}(k)$ in our case) 
in our 21~cm signal modeling (Eqs.~\ref{eq:EEps}, \ref{eq:TTps} and \ref{eq:TEps}).

In order to calculate the temperature and polarization anisotropy signals, 
we need relevant power spectra \emph{on all scales}.
To this end, it is useful to study the bias parameter, 
$b^2_{\rm HI}(k)\equiv P_{\rm HI}(k)/P_\delta(k)$, and $b_{\rm HI,\times} \equiv P_{\rm HI\,\delta}(k)/P_\delta(k)$. 
We present in Figure~\ref{fig:HIbias} the simulated H~{\small I} bias, $b^2_{\rm HI}$, 
for all the reionization models in Table~\ref{tab:EoRparam}. (The bias from the cross-power $b_{\rm HI,\times}$ agrees with $b_{\rm HI}(k)$ on large scales.) 
For the $k$ range covered by simulations, 
we stitch together the results from the large and small boxes 
and discard data from $k>2$ Mpc$^{-1}$, to avoid the alias effect. 
Figure~\ref{fig:HIbias} shows that at a given global ionized fraction, 
the H~{\small I} bias, $b^2_{\rm HI}(k)$, is almost the same for all these models, 
regardless of the reionization parameters.
As for its scale-dependence,
since it is expected that the spatial distribution of the H~{\small I} gas follows the underlying total matter on large scales, the H~{\small I} bias should be constant on super-bubble scales throughout the EoR. Figure~\ref{fig:HIbias} confirms this point.\footnote{A caveat is the bias at the end of the EoR 
$(\bar Q_{\rm H\,{\tiny II}} \simeq 0.99)$, as shown in the leftmost panel of Figure~\ref{fig:HIbias}, showing a scale-dependence. This is likely a numerical artifact which arises because the seminumerical reionization simulation based on the excursion set model breaks down at that time.} 
To accommodate the simulation results at all redshifts,
we adopt a power-law extrapolation for the H~{\small I} bias, $b^2_{\rm HI}(k)\propto k^n$, 
on scales beyond either side of the simulation coverage.
In contrast, the large-scale behavior of the H~{\small I} bias from the toy model is very different, 
because its $P_x(k)$ approaches constant as $k\to0$ in Eq.~(\ref{eq:ZFH04model}), resulting in a red-tilted $b^2_{\rm HI}(k)$.

\section{Results} \label{sec:results}

We discuss the patterns and physical implications
of the 21~cm TT, TE and EE power spectra with a fiducial model in \S\ref{ssec:aps}, 
and focus on the evolution of $C^{\rm TE}_\ell$ and $C^{\rm EE}_\ell$ during the EoR
and present their dependence on the global ionization history in \S\ref{ssec:apsevo}. 

\subsection{Generic feature of angular power spectra} \label{ssec:aps}

\begin{figure}
\resizebox{\columnwidth}{!}{\includegraphics{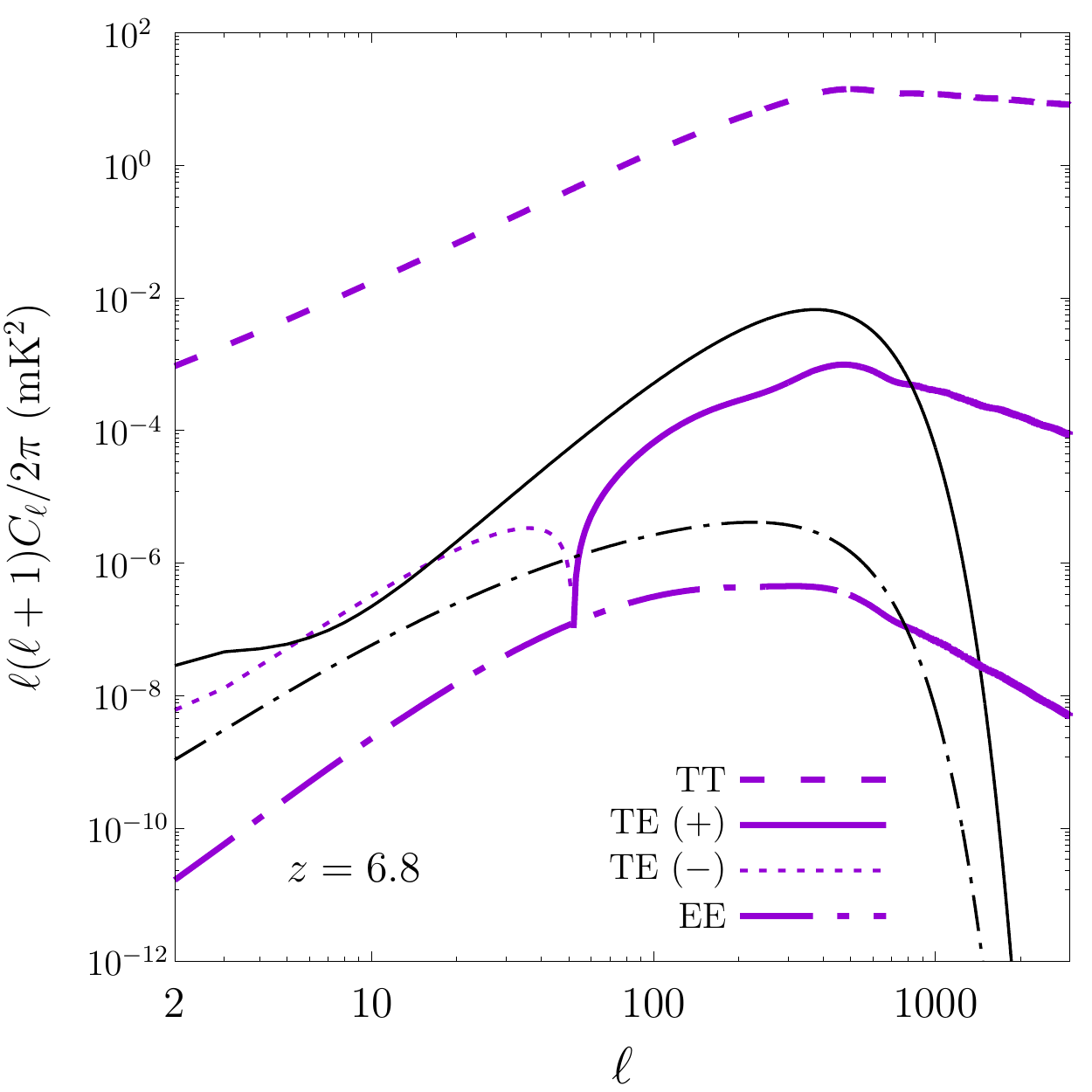}}
\caption{Angular power spectra of the 21~cm temperature (TT, dashed), 
$E$-mode polarization (EE, dot-dashed) and the temperature-polarization cross-correlation (TE, solid for positive part and dotted for negative part) at $z_{\rm em}=6.8$ ($\bar Q_{\rm H {\tiny II}}\approx 0.82$). Shown are the results from the Model 1 of seminumerical simulation (purple lines), and from the toy model (Eq.~\ref{eq:ZFH04model}, thin black lines). The TE power spectrum from simulation is plotted with its absolute value, with the dotted/solid line type marking the negative/positive part on the left/right-hand side of a sharp zero-crossing. 
}
\label{fig:angularps1}
\end{figure}

While all the three models in Table \ref{tab:EoRparam} take reasonable reionization parameters 
and satisfy all current observational constraints, we choose Model 1 as the fiducial model just for illustration purpose.
In Figure~\ref{fig:angularps1}, we present its 21~cm temperature and polarization angular power spectrum
at $z_{\rm em}=6.8$ ($\bar Q_{\rm H {\tiny II}}\approx 0.82$). 
It shows that the temperature-polarization cross-power spectrum, $C^{\rm TE}_\ell$, is about four orders of magnitude smaller than the temperature power spectrum $C^{\rm TT}_\ell$, and the polarization power spectrum $C^{\rm EE}_\ell$ is three orders of magnitude even smaller than $C^{\rm TE}_\ell$. This will make the detection of 21~cm polarization signal very difficult, as we will see in \S\ref{sec:detection}. 
Regarding the comparison with the toy model, on the other hand, 
it shows that the toy model always underestimates the polarization (EE and TE) powers on small scales but overestimates them on large scales. 

The EE power spectrum $C^{\rm EE}_\ell$ in Figure~\ref{fig:angularps1} shows a plateau on the intermediate angular scales and turns over at the scale of the typical bubble size ($\ell\sim s/R$).
The bump feature is caused by
the turnover in the H~{\small I} density power spectrum (see Figure~\ref{fig:Pk}).
This feature due to the morphological structure of cosmic H~{\small II} regions during the EoR is also reflected in $C^{\rm TE}_\ell$ at the angular scale of bubble size ($\ell\sim s/R$), as shown in Figure~\ref{fig:angularps1}.

\begin{figure}
\resizebox{\columnwidth}{!}{\includegraphics{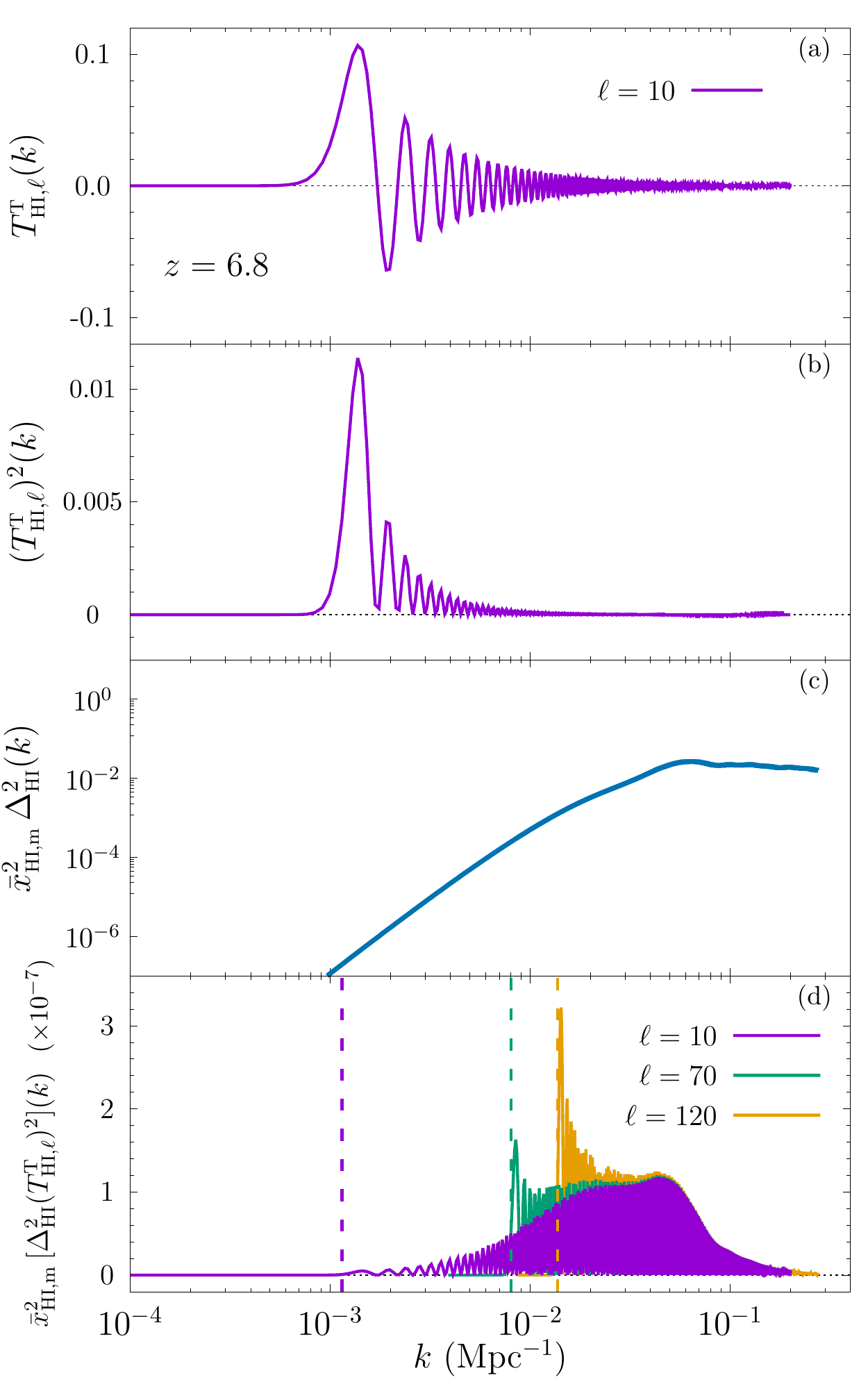}}
\caption{Contributions to the $C^{\rm TT}_\ell$ integration at $z_{\rm em}=6.8$ in Model 1.  
(a) temperature transfer function of H~{\scriptsize I} density fluctuations, $T^{\rm T}_{{\rm HI},\ell}(k)$ with $\ell=10$. 
(b) $\left(T^{\rm T}_{{\rm HI},\ell}\right)^2(k)$ with $\ell=10$. 
(c) the dimensionless H~{\scriptsize I} power spectrum, $\Delta^2_{\rm HI}(k)\equiv k^3P_{\rm HI}(k)/2\pi^2$, weighted with relevant factor, from simulation. (Note that the power spectra at both high and low ends of the $k$ range beyond the simulation coverage is obtained by extrapolating the H~{\scriptsize I} bias from simulation with power-law fits.)  
(d) the leading term in the integrand of $C^{\rm TT}_\ell$ in Eq.~(\ref{eq:TTps}), 
$\bar x_{\rm HI,m}^2\,\Delta^2_{\rm HI}\left(T^{\rm T}_{{\rm HI},\ell}\right)^2$ (i.e.\ product of the quantities in (b) and (c)), as a function of wavenumber $k$, 
at various angular scales, $\ell=10/70/120$ (purple/green/yellow). The characteristic scale $k_* = \ell/s$ at $z_{\rm em}=6.8$ is indicated by the vertical lines for $\ell$ of the same color.
}
\label{fig:transferT}
\end{figure}

\begin{figure}
\resizebox{\columnwidth}{!}{\includegraphics{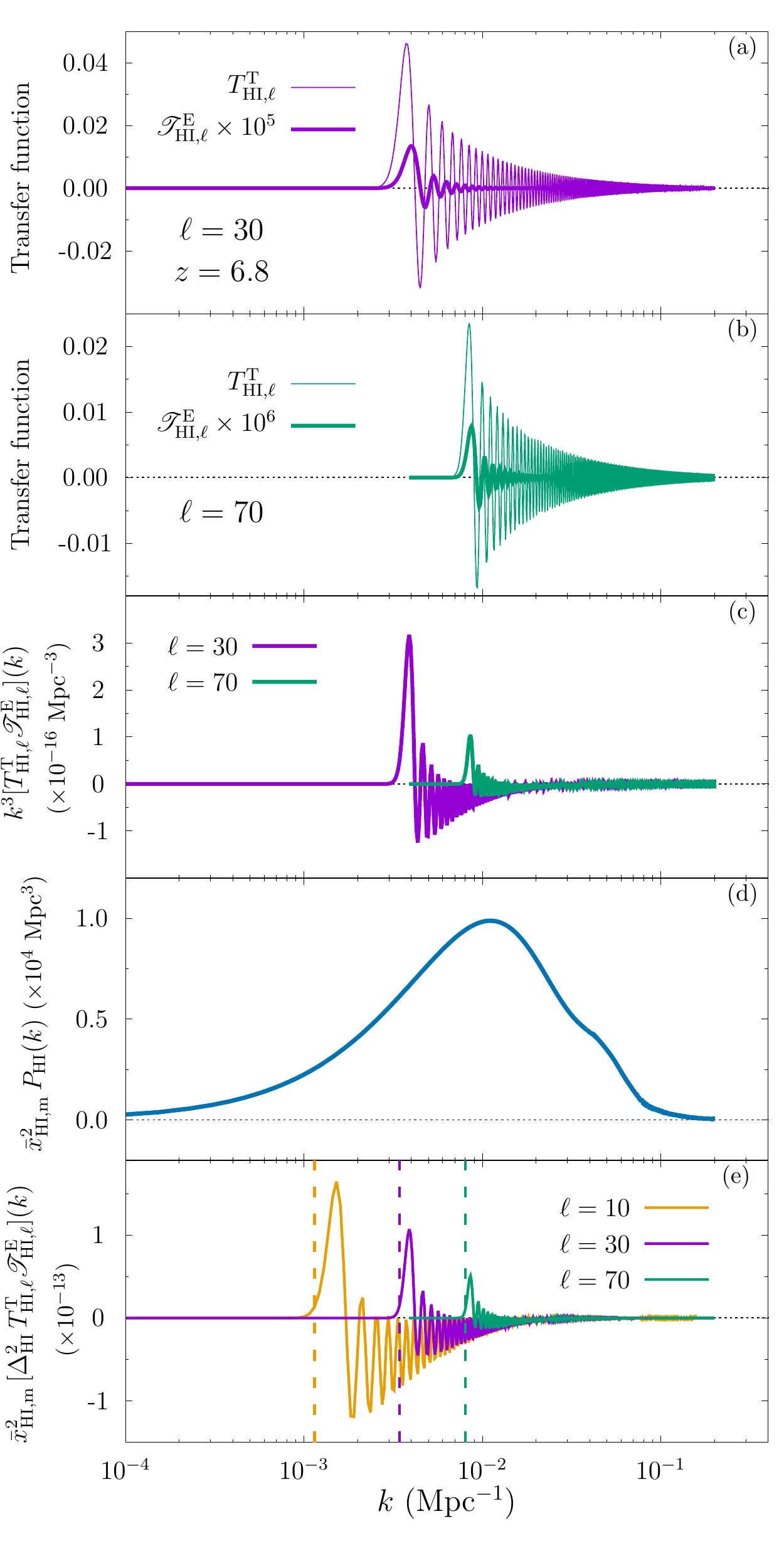}}
\caption{Contributions to the $C^{\rm TE}_\ell$ integration at $z_{\rm em}=6.8$ in Model 1. 
(a) temperature and polarization transfer functions of H~{\scriptsize I} density fluctuations 
$T^{\rm T}_{{\rm HI},\ell}(k)$ (thin) and $\mathscr{T}^{\rm E}_{{\rm HI},\ell}(k)$ (thick) at $\ell=30$. 
(b) same as in (a) but for $\ell=70$. 
(c) $k^3 T^{\rm T}_{{\rm HI},\ell} \mathscr{T}^{\rm E}_{{\rm HI},\ell}$ for $\ell=30/70$ (purple/green). 
(d) H~{\scriptsize I} density power spectrum $P_{\rm HI}(k)$, multiplied by $\bar x^2_{\rm \scriptsize HI,m}$ at this redshift.
(e) the leading term in the integrand of $C^{\rm TE}_\ell$ in Eq.~(\ref{eq:TEps}), 
$\bar x_{\rm HI,m}^2\,\Delta^2_{\rm HI}T^{\rm T}_{{\rm HI},\ell} \mathscr{T}^{\rm E}_{{\rm HI},\ell}$ as a function of wavenumber $k$, at various angular scales, $\ell=10/30/70$ (yellow/purple/green). The characteristic scale $k_* = \ell/s$ at $z_{\rm em}=6.8$ is indicated by the vertical lines for $\ell$ of the same color.}
\label{fig:transferE}
\end{figure}

\begin{figure*}
\resizebox{\textwidth}{!}{\includegraphics{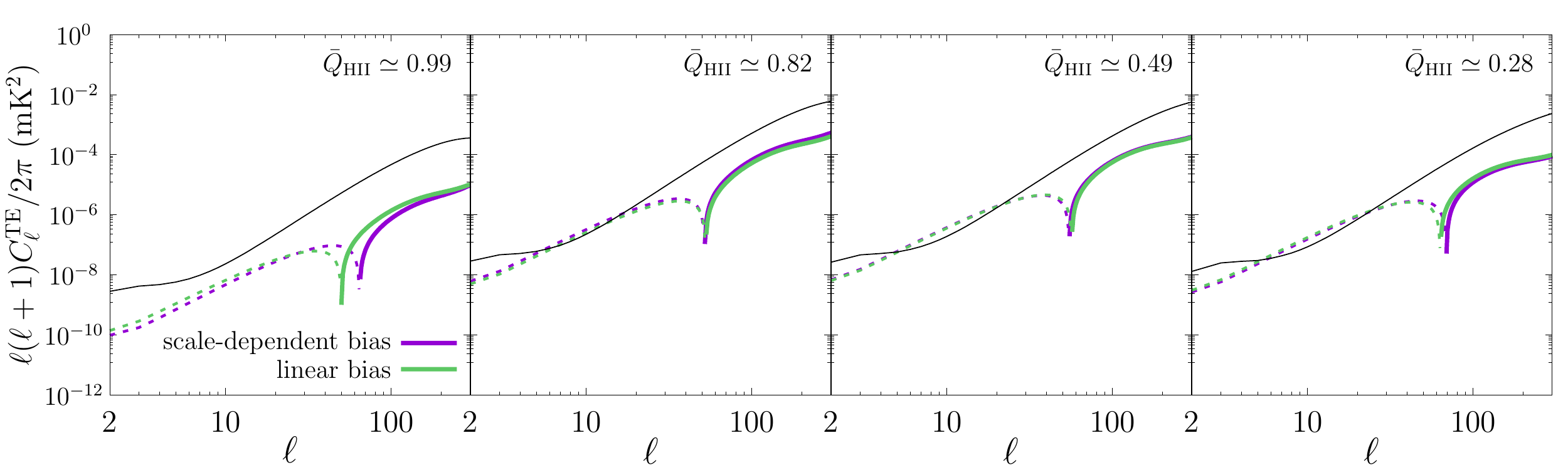}}
\caption{Large-scale TE angular power spectra at different ionized fractions during the EoR in seminumerical simulation Model 1 (colored lines) and in the toy model (Eq.~\ref{eq:ZFH04model}, thin black lines). 
$C^{\rm TE}_\ell$ is plotted with its absolute value, with the dotted/solid line type marking the negative/positive part on the left/right-hand side of a sharp zero-crossing.
For seminumerical simulations, shown are the results by employing a scale-dependent (power-law) extrapolation of the H~{\scriptsize I} bias on large scales (purple lines), and by assuming a constant bias on large scales (green lines), respectively.
}
\label{fig:TEps_model_1}
\end{figure*}

On large scales, the temperature power spectrum turns out to be like a white noise, 
i.e.\ $C^{\rm TT}_\ell=$ constant for small $\ell$ \citep{2007PhRvD..76h3005L}. This is 
because for small $\ell$ it is dominated by the contributions from the $k$-modes of small scales ($ks>\ell$) rather than those of comparable scales.
This is illustrated in Figure~\ref{fig:transferT}. 
While the temperature transfer function $T^{\rm T}_{{\rm HI},\ell}(k)$ peaks near $k_* \simeq\ell/s$, 
the H~{\small I} power spectrum peaks at much smaller scale. 
As a result, it turns out that the integrand of $C^{\rm TT}_\ell$ (see Eq.~\ref{eq:TTps}) peaks at a scale much smaller than $k_*$ for small $\ell$ (see the bottom panel of Figure~\ref{fig:transferT}). 
Given the shape of $\Delta^2_{\rm HI}(k)$, only for large enough $\ell$ ($\ell \gtrsim 100$) 
does the majority of contributions to the $C^{\rm TT}_\ell$ integration 
come from the modes of comparable scales $k_*$.

However, this is not the case for the large-scale behavior of the temperature-polarization cross-power spectrum, 
$C^{\rm TE}_\ell$, which involves the transfer function of the $E$-mode polarization.
Figure~\ref{fig:transferE} shows that in $C^{\rm TE}_\ell$ the integration in Eq.~(\ref{eq:TEps}) is always dominated by the modes near the characteristic scale $k_* = \ell/s$, for all $\ell$. 

Furthermore, since $T^{\rm T}_{{\rm HI},\ell} \mathscr{T}^{\rm E}_{{\rm HI},\ell}$
can be negative, as shown in Figure~\ref{fig:transferE}, 
the integrand of $C^{\rm TE}_\ell$ has negative parts, and those negative parts become more important as $\ell$ is smaller.
As a result, the integration over Fourier modes can result in negative values when $\ell$ is small. 
This is why the $C^{\rm TE}_\ell$ from simulation results shows a zero-crossing at $\ell < 100$, 
as we find in Figure~\ref{fig:angularps1}. 
However, the same figure also shows that $C^{\rm TE}_\ell$ from the toy model is always positive. 
This is because the toy model significantly overestimates the H~{\small I} power spectrum at small $k$ 
where $T^{\rm T}_{{\rm HI},\ell}$ and $\mathscr{T}^{\rm E}_{{\rm HI},\ell}$ are mostly positive, 
thereby enhancing the values of the integrand toward small $k$. 
This comparison between simulation and toy model results implies that the sign of $C^{\rm TE}_\ell$ is sensitive to the H~{\small I} bias at large scales. 

Specifically, we explore the impact of the large-scale scale-dependence of the H~{\small I} bias on the zero-crossing angular scale of $C^{\rm TE}_\ell$. For this purpose, we consider to extrapolate the H~{\small I} bias at large scales as a constant, instead of a power-law dependence. This difference is particularly significant for $\bar Q_{\rm H {\tiny II}}\simeq 0.99$ (see the leftmost panel of Figure~\ref{fig:HIbias}). 
In Figure \ref{fig:TEps_model_1}, we find that, for constant bias extrapolation, the zero-crossing angular scale evolves to larger angular scales as reionization proceeds, from $\ell\sim 70$ at $\bar Q_{\rm H {\tiny II}}\simeq 0.28$ to $\ell\sim 50$ at $\bar Q_{\rm H {\tiny II}}\simeq 0.99$. 
Meanwhile, for a given $\bar Q_{\rm H {\tiny II}}$, 
the change of the scale-dependence of the large-scale H~{\small I} bias results in a shift of the zero-crossing angular scale. 
The explanation is that a linear H~{\small I} bias effectively modulates the integrand with more weights on large scales than the bias with a positive power-law extrapolation,  
thus shifting the zero-crossing to a larger angular scale. 
Therefore, the zero-crossing feature in $C^{\rm TE}_\ell$ may be used to probe the \emph{scale-dependence} of the H~{\small I} bias during the EoR on super-bubble scales.

\subsection{Evolution and model-dependence} \label{ssec:apsevo}

\begin{figure}
\resizebox{\columnwidth}{!}{\includegraphics{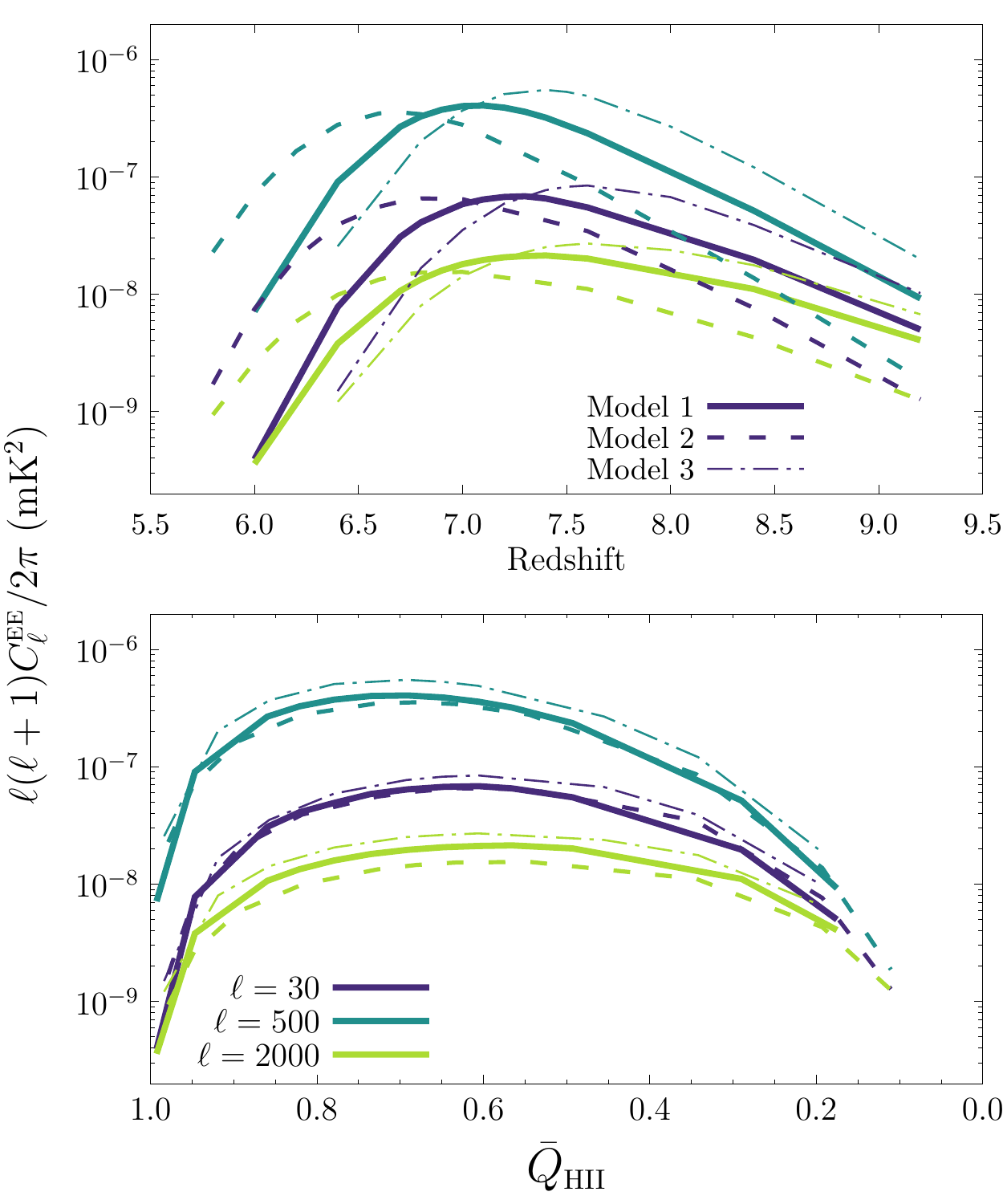}}
\caption{Evolution of the EE power spectrum $C^{\rm EE}_\ell$ at different angular scales $\ell=30/500/2000$ (purple/dark green/light green) during the EoR for different Model 1/2/3 (solid/dashed/dot-dashed line). 
(Top) $C^{\rm EE}_\ell$ as a function of redshift.
(Bottom) $C^{\rm EE}_\ell$ as a function of the global H~{\scriptsize II} volume filling fraction.}
\label{fig:evolution_EE}
\end{figure}

\begin{figure}
\resizebox{\columnwidth}{!}{\includegraphics{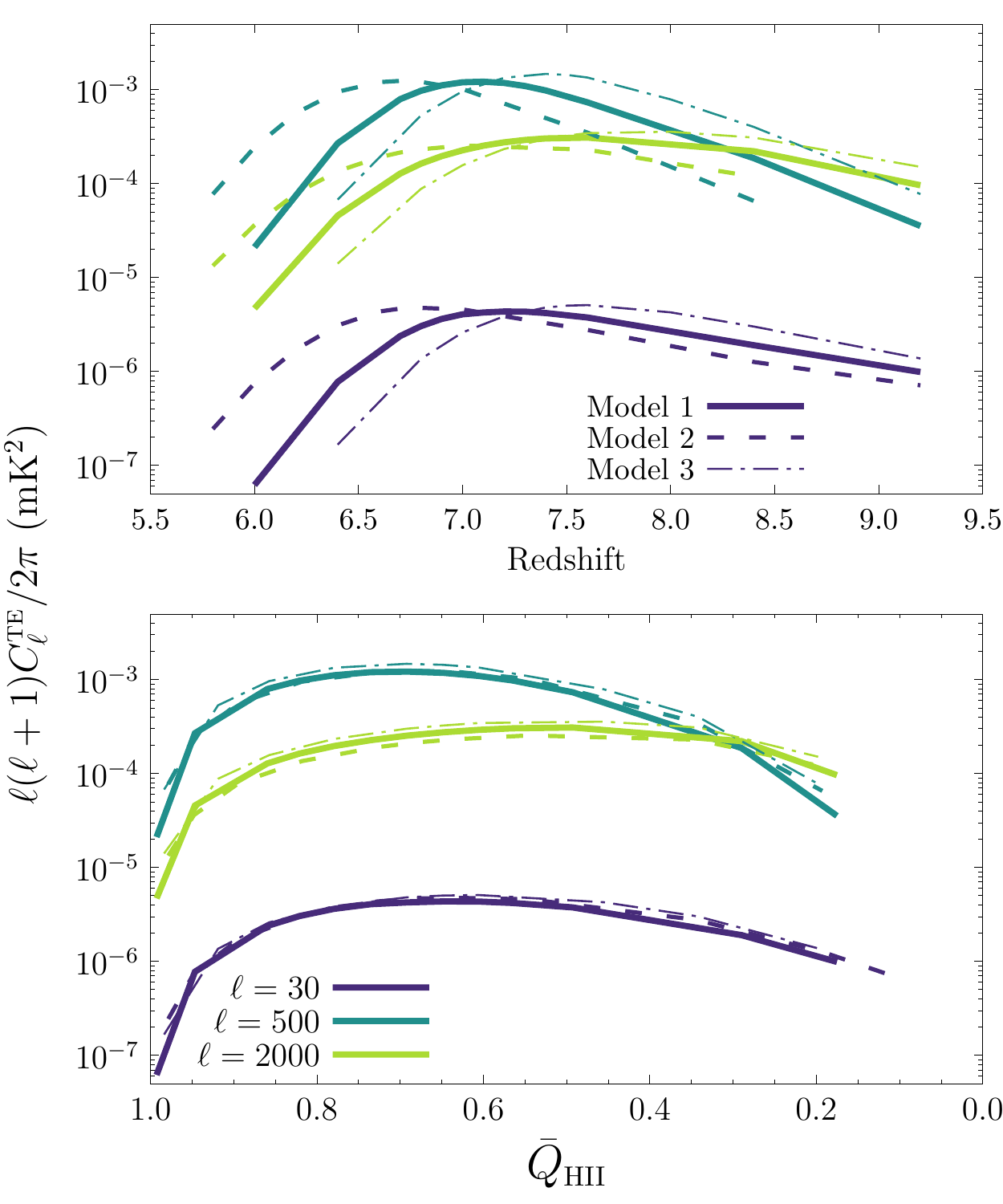}}
\caption{Same as Figure~\ref{fig:evolution_EE} but for the TE power spectrum $C^{\rm TE}_\ell$. 
The results for $\ell=30$ are plotted with absolute values of $C^{\rm TE}_\ell$ since they are negative.}
\label{fig:evolution_TE}
\end{figure}

\begin{figure}
\resizebox{\columnwidth}{!}{\includegraphics{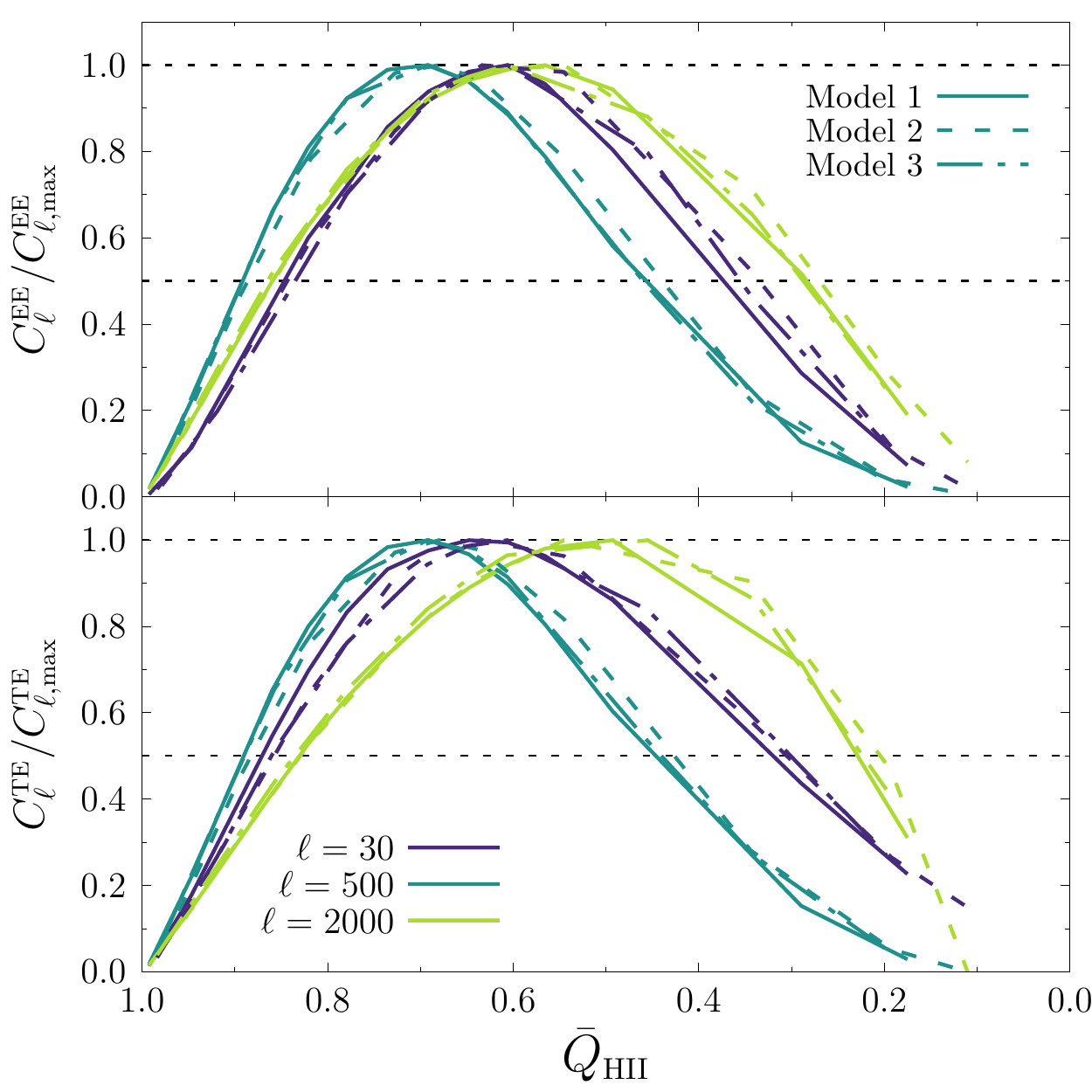}}
\caption{The ratio $C^{\rm EE}_{\ell}/C^{\rm EE}_{\ell,{\rm max}}$ (top) and $C^{\rm TE}_\ell / C^{\rm TE}_{\ell,{\rm max}} $ (bottom) as a function of the global H~{\scriptsize II} volume filling fraction, for different Model 1/2/3 (solid/dashed/dot-dashed line) 
at different angular scales $\ell=30/500/2000$ (purple/dark green/light green). 
In the bottom panel, the ratios at $\ell=30$ are plotted with absolute values of $C^{\rm TE}_\ell$ since they are negative.}
\label{fig:linear_evo}
\end{figure}

We plot the evolution of EE and TE power spectra, 
in Figures~\ref{fig:evolution_EE} and \ref{fig:evolution_TE} respectively, 
at three representative angular scales corresponding to linear scales ($\ell=30$), 
the plateau (or the peak, $\ell=500$), and sub-bubble scales ($\ell=2000$).
For both $C^{\rm EE}_\ell$ and $C^{\rm TE}_\ell$, 
all curves exhibit a similar trend in the evolution of the amplitude of the signal: 
as reionization proceeds, the signal first rises to its peak 
and then gradually drops to almost zero when the IGM becomes fully ionized.
The location of the peak is dependent on $\ell$, 
varying with a broad range $\bar Q_{\rm H {\tiny II}}\sim 0.50 - 0.70$.
The top panels of Figures~\ref{fig:evolution_EE} and \ref{fig:evolution_TE} 
show that the powers at a given redshift are highly model-dependent. 
However, 
if we compare the powers \emph{at a given global ionized fraction} $\bar Q_{\rm H {\tiny II}}$
from different models, we find that both the EE and TE powers 
display better convergence across these models in terms of their evolution trends (with small variations only).

Inspired by this finding, we plot the ratio $C^{\rm EE}_{\ell}/C^{\rm EE}_{\ell,{\rm max}}$ and $C^{\rm TE}_\ell / C^{\rm TE}_{\ell,{\rm max}} $ as function of $\bar Q_{\rm H {\tiny II}}$ in Figure~\ref{fig:linear_evo}, 
where $C^{\rm EE}_{\ell,{\rm max}}$ ($C^{\rm TE}_{\ell,{\rm max}}$) is the maximum value of $C^{\rm EE}_\ell$ ($C^{\rm TE}_\ell$) during the EoR at a given $\ell$.
We find that the relations of these ratios and $\bar Q_{\rm H {\tiny II}}$ indeed display even better model-independence, for all three representative angular scales $\ell$. 
If 21~cm polarization measurement becomes technically achievable, then the robust mapping between the polarization signal and the global ionized fraction can be exploited to infer $\bar Q_{\rm H {\tiny II}}$ from the $C^{\rm EE}_{\ell}$ or $C^{\rm TE}_\ell$ measurements.
This will help extract at least some part of the reionization history in later stages of the EoR.

Besides, Figure~\ref{fig:linear_evo} shows that the full width at half maximum (FWHM) of the signal over the EoR is also insensitive to the ionization history, e.g.\ the FWHM of the power at $\ell=500$ approximately spans the range from $\bar Q_{\rm H {\tiny II}}=0.50$ to $0.90$. 
In other words, once 21~cm polarization measurement becomes feasible, 
the polarization signal (analyzed using $C^{\rm TE}_\ell$ and $C^{\rm EE}_\ell$) from the EoR may be observed over a broad redshift (frequency) range --- $\Delta z\sim 1$ ($\Delta\nu_{\rm obs}\sim30$ MHz) in terms of the FWHM --- regardless of the global ionization history. 

In short summary, our results show that 
the polarization of redshifted 21~cm lines can potentially 
probe large-scale spatial fluctuations of neutral gas in the IGM,
as well as the late-stage ionization history.
In the next section we will discuss the prospects of its detection
by upcoming intensity mapping experiments like SKA1-low.

\begin{figure*}
\resizebox{\textwidth}{!}{\includegraphics{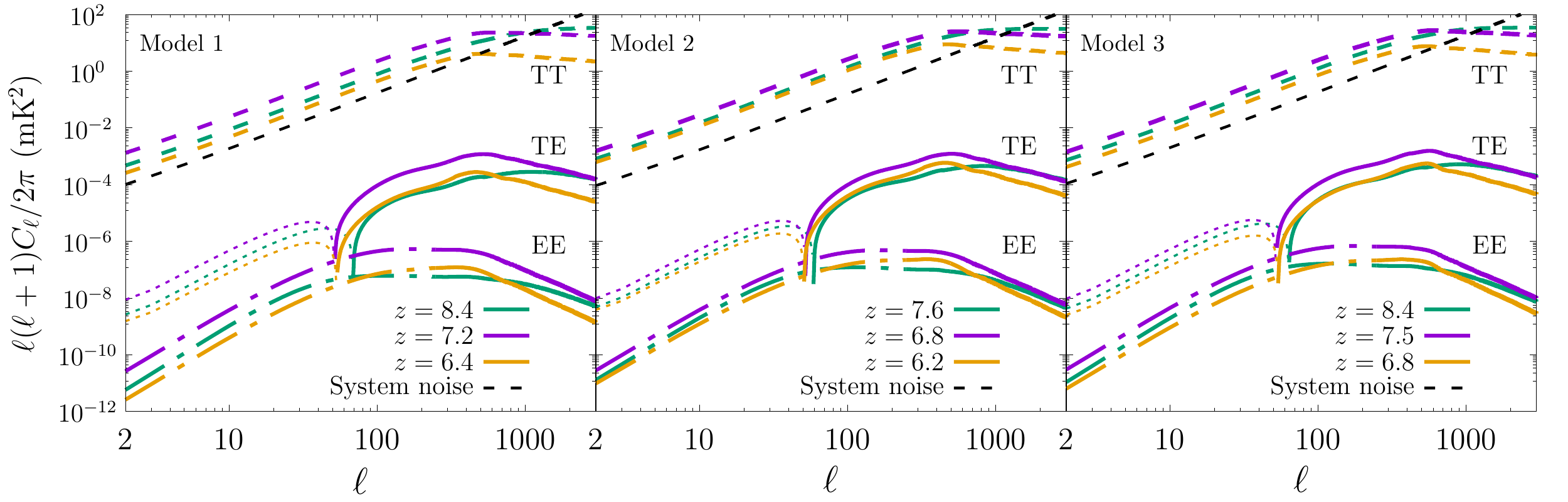}}
\caption{Angular power spectra of the 21~cm temperature (TT, dashed), 
$E$-mode polarization (EE, dot-dashed) and the temperature-polarization cross-correlation (TE, solid/dotted for the positive/negative part, respectively) from different Model 1/2/3 (left/middle/right panels). For each model (with different ionization history), shown are the results at various redshifts corresponding to $\bar Q_{\rm H {\tiny II}}\approx ~0.35$ (green), $0.65$ (purple) and $0.95$ (yellow). 
TE power spectra are plotted with their absolute values,  with the dotted/solid line type marking the negative/positive part on the left/right-hand side of a sharp zero-crossing. 
The black dashed line represents an estimate of the system noise of an SKA1-low survey
with total observing time of 4096 hrs. The frequency band of the noise power spectrum 
corresponds to the redshift of the largest signal shown in each panel.
}
\label{fig:angularps_all}
\end{figure*}

\section{Detection Prospects}  \label{sec:detection}

While the temperature signal has hitherto been the focus of 21~cm cosmology, 
all Stokes parameters are actually often measured in intensity mapping surveys
with low-frequency coverage \citep[e.g.,][]{2019A&A...622A...1S, 2020PASA...37...29R}.
In fact, polarization components have long been sought
with a variety of science goals \citep[see][for a brief review]{2020Galax...8...53H},
or for purposes of calibration and understanding instrumental systematics
\citep[e.g.,][]{2017PASA...34...40L, 2018MNRAS.478.1484G}.

Nevertheless, a practical measurement of 
the cosmological polarization signal from the EoR 
may be too ambitious a goal for current-generation experiments, 
due to several major challenges in 21~cm polarimetry --- 
calibration \citep[e.g.,][]{1996A&AS..117..149S,2013ApJ...771..105B,2019ApJ...882...58K},
foreground contamination from polarized diffuse synchrotron emission \citep{2007ApJ...665..355K,2016ApJ...830...38L,2019A&A...623A..71V}, 
instrumental leakage of intensity into polarization \citep[e.g.,][]{2018MNRAS.476.3051A},
the Faraday rotation effect which rotates the linear polarization 
by intervening magnetic fields of cosmic origins or in the ionosphere of the Earth \citep[e.g.,][]{2010MNRAS.409.1647J, 2011A&A...527A.107S, 2013ApJ...769..154M},
and depolarization effects \citep{1966MNRAS.133...67B, 2020ApJ...894...38P}.

Concerning these uncertainties, 
there have been continuous progresses in controlling the instrumental systematics  
and correcting for ionospheric activities to good levels,
whereas the mitigation of radio foregrounds still seems a formidable task.
For polarization, foreground mitigation is even more difficult than that for intensity,
since both foregrounds and the 21~cm polarization signals are additionally subject to Faraday rotation, 
with different Faraday depths along the LoS.
Foreground subtraction algorithms, e.g.\ $m$-mode formalism \citep{2014ApJ...781...57S, 2015PhRvD..91h3514S}, and pseudo-$C_{\ell}$ algorithm \citep{2019MNRAS.484.4127A}, as well as the foreground avoidance strategy \citep[e.g.,][]{2010ApJ...724..526D, 2012ApJ...752..137M,2014PhRvD..90b3018L}, 
may provide efficient techniques for foreground mitigation. 
However, even if polarized foregrounds could be cleaned perfectly,
Faraday rotation by the Galactic magnetic field alone can 
threaten the reconstruction of the cosmological linear polarization signal \citep{2014PhRvD..89l3002D}.

On the bright side, Faraday rotation provides an avenue to study cosmic magnetic fields. 
By broadband observations and rotation measure (RM) synthesis techniques, 
modern radio astronomy aims at producing high precision RM maps 
which resolve structure of magnetic fields along each LoS \citep{2015ASSL..407...41H}.
As an example, an RM grid survey is planned on the SKA1-mid telescope \citep{2015aska.confE..92J}.
Future magnetism science might then offer 
new approaches to separating polarized foregrounds 
and reconstructing the full Stokes parameters of the cosmological signal at the same time. 
Such a possibility is worth further scrutiny.

In this paper, we only demonstrate the detection sensitivity
by the conventional estimate of the measurement noise---the rms noise fluctuation 
in the visibility of a fixed baseline in one frequency channel centered at $\nu_{\rm obs,0}$
\citep{2004tra..book.....R},
\begin{equation}
	\Delta V^{\rm N}=\frac{T_{\rm sys}}{A_{\rm eff}}\frac{\lambda_{\rm obs,0}^2}{\sqrt{\Delta\nu_{\rm obs,0}\,t_v}},
\end{equation}
where $T_{\rm sys}(\nu_{\rm obs,0})$ is the system temperature, 
which is the sum of the sky brightness temperature and the receiver temperature,
$A_{\rm eff}$ is the effective collection area of one antenna/station,
$\Delta\nu_{\rm obs,0}$ is the bandwidth of the channel
and $t_v$ is the observation time in this channel.
$T_{\rm sys}/A_{\rm eff}$ characterizes the sensitivity of an antenna.
In the case of SKA1-low, baselines are formed by pairs of stations and 
the zenith sensitivity within $45^{\circ}$ for one station is 
$A_{\rm eff}/T_{\rm sys}\approx 1.5~{\rm m}^2/{\rm K}$
across relevant frequencies. Our assumption of sensitivity follows the SKA1-low antenna selection book\footnote{See the link at \href{https://www.skatelescope.org/wp-content/uploads/2020/06/SKA1_LOW_Antenna_Selection_Aug2017.pdf}{SKA1\_LOW\_Antenna\_Selection.pdf}.}.

For illustration, we consider a drift-scan survey with maximal sky coverage by SKA1-low. 
Ignoring details of the map-making process in practice,
we estimate the measurement uncertainty of $C_\ell$ 
by the angular power spectrum of $\Delta V^{\rm N}$ \citep{2004ApJ...608..622Z, 2020MNRAS.494.4043M}, 
\begin{equation} \label{eq:noiseCl}
	C_\ell^{\rm N}\approx \frac{(2\pi)^2 (T_{\rm sys}/A_{\rm eff})^2}{\Delta\nu_{\rm obs,0}\,t_0}
	\frac{\lambda^2_{\rm obs,0}\,A_{\rm total}}{2N_{\rm p}N^2_{\rm s}},
\end{equation} 
where $t_0$ is the total observing time, $N_{\rm p}=2$ is the number of polarization states,
$A_{\rm total}$ is the total area covered by the array 
and $N_{\rm s}$ is the number of stations within the array.
Eq.~(\ref{eq:noiseCl}) assumes a roughly uniform coverage of Fourier space
during the course of the survey, preserving a white noise. 
For this reason, we only take into account the ``core'' of the actual SKA1-low array
within the radius of about $500$ meters, inside which 224 stations are closely packed\footnote{These parameters are adopted from the recent revision of the SKA1 baseline design document
(SKA-TEL-SKO-0001075).}.
We take the bandwidth to be $\Delta\nu_{\rm obs,0}=0.13$ MHz 
(2304 bands in total within the $50-350$ MHz full SKA1-low bandwidth)
and consider a total observing time of 4096 hrs
(e.g., 512 nights with 8hrs per night).

The resulted noise power spectra are shown (labeled as ``System noise'') in Figure~\ref{fig:angularps_all}. 
For comparison, we also plot the angular power spectra from different ionization histories. For each ionization history, the output redshifts
are chosen such that they nearly correspond to
$\bar Q_{\rm H {\tiny II}}\approx 0.35$, $0.65$ and $0.95$. 
Although the error estimate here is primitive,
the conclusion for $C_\ell^{\rm TT}$ is similar to that from \citet{2020MNRAS.494.4043M} 
who accounted for the LC effect, that given reasonable integration time 
SKA1-low may detect the temperature signal by a signal-to-noise ratio (SNR) of over 10.
However, for $C_\ell^{\rm TE}$ and $C_\ell^{\rm EE}$, 
the 21~cm polarization signal is still several orders of magnitude below the noise power. 
This conclusion holds even if we relax the total observation time to $t_0=20000$ hrs under the current design sensitivity of SKA1-low. 
It contrasts with the optimistic conclusion in BL05 because 
(1) the polarization signal from the EoR is overestimated at large scales in BL05, 
and (2) current design sensitivity of SKA1-low is much worse than the early illustration in BL05.
Therefore, with the designed sensitivity of the current-generation interferometer array experiment, 
it is not feasible to measure the 21~cm polarization power spectrum.

\section{Summary} \label{sec:conclusions}

In this paper we reexamine the possibility of using the redshifted 21~cm polarization signal 
to probe cosmic reionization, in light of the upcoming H~{\small I} intensity mapping surveys. 
For linearly polarized 21~cm lines due to Thomson scattering -- the dominant mechanism for the 21~cm polarization signal -- 
we improve the prediction of all-sky angular power spectra of the polarization autocorrelation and the temperature-polarization cross-correlation upon the previous work (BL05). 
We take into account the nonlinear effects due to inhomogeneous reionization, 
and perform realistic modeling of patchy reionization with seminumerical simulations. 

We find that both power spectra $C^{\rm EE}_\ell$ and $C^{\rm TE}_\ell$ are enhanced on the sub-bubble (i.e.~small) scales, compared to the previous predictions from BL05. 
On the large scales, however, the peak amplitudes of both power spectra, which correspond to the characteristic scale of bubbles, are smaller than the BL05 results.
This is partly due to the fact that our formalism includes the anti-correlation between the matter density field and the neutral fraction field. 
In particular, for all the global ionization histories considered herein, 
$\sqrt{\ell (\ell+1)\,C^{\rm EE}_{\ell, \rm max}/2\pi}$ can only reach $\sim 1\,\mu$K 
and $\sqrt{\ell (\ell+1)\,C^{\rm TE}_{\ell, \rm max}/2\pi} \sim 0.03$ mK. 

For the temperature-polarization cross-power spectrum, 
we find that $C^{\rm TE}_\ell$ displays a zero-crossing at $\ell<100$, 
and its angular scale is sensitive to the scale-dependence of the large-scale H~{\small I} bias during the EoR.

We demonstrate that the correlation between $C_\ell/C_{\ell, \rm max}$ and the global ionized fraction $\bar Q_{\rm H {\tiny II}}$, for both EE and TE power spectra, is robust against the variation of reionization parameters. 
This correlation may be exploited to infer the global ionized fraction from the measurement of $C_\ell/C_{\ell, \rm max}$, 
so as to reconstruct at least some part of the ionization history from the midpoint to the late stages of the EoR.

Regarding the detectability of the 21~cm polarization signal, the thermal noise, foregrounds and systematics remain challenges to the observations. 
Even with Faraday rotation corrected and foreground removed, the thermal noise in the polarization alone is still much larger than the signal in our new forecast, with the sensitivity of the SKA1-low telescope within reasonable integration time. 


There still remains the possibility to probe the 21~cm polarization signal 
by its cross-correlation with other cosmic tracers (see \citealt{2015aska.confE...8J} and references therein).
For example, the cross-correlation of 21~cm polarization with the secondary CMB polarization due to patchy reionization
\citep[][]{2007PhRvD..76d3002D} might be worth consideration 
since they are both generated by Thomson scattering off free electrons. Similar cross-correlation technique has been explored between 21~cm temperature and CMB kSZ signals
\citep[e.g.,][]{2004PhRvD..70f3509C,2006ApJ...647..840A}.

\acknowledgments

This work is supported by National SKA Program of China (Grant No.~2020SKA0110401), NSFC (Grant No.~11821303), and National Key R\&D Program of China (Grant No.~2018YFA0404502). 
We thank Shifan Zuo and Paulo Montero-Camacho for valuable comments and discussions, and thank the anonymous referee for constructive suggestions.
Numerical evaluations and simulations in this work were ran at the Venus cluster at the Tsinghua University.

%






\appendix

\section{The 21~cm basics: relativistic formalism} \label{app:deltaTb}

This appendix contains the full relativistic formalism for the 21~cm signal from the EoR
which we omit in \S\ref{ssec:deltaTb} and leads to the starting point of our calculation, 
Eqs.~(\ref{eq:tbpostheating}) and (\ref{eq:tbkfinal}).
It closely follows those in \citet{2013PhRvD..87f4026H}.

For the interest of this paper, the concept of observers extends to fictitious ones such as free electrons on the photon path.
We consider such ``observers'' on the past light cone of the present-day Earth labeled by their {\it relative redshift} $z_{\rm obs}$ between the ``observers'' and the Earth today, 
$1+z_{\rm obs}=\nu_{\rm obs}/\nu_{\rm obs,0}$, where $\nu_{\rm obs}$ and $\nu_{\rm obs,0}$ 
are the radiation frequencies seen by the ``observer'' and on the Earth today, respectively. 
For a fixed observer, the location of 21~cm emission events 
is labeled by FLRW coordinates $(t_{\rm em}, \vec x+\vec r)$.
$\vec x$ is the location of the observer and $\vec r\equiv-r\hat n$.

The fundamental observable in 21~cm radiation for an ``observer'' 
is the differential 21~cm brightness temperature along a line of sight (LoS)
per observed frequency bin, 
$\delta T_{\rm b}(\nu_{\rm obs}, \hat n)$, generally written as
\begin{equation}\label{eq:deltatb}
	\delta T_{\rm b}(\nu_{\rm obs}, \hat n)=\left[\frac{T^g_{\rm s}}{1+z}-T^{\rm obs}_{\rm CMB}(\hat n)\right]\,(1-e^{-\tau_{\nu_{\rm obs}}}),
\end{equation}
where $T^g_{\rm s}$ is the spin temperature of the H~{\small I} gas\footnote{In this paper, 
the superscript `g' denotes quantities in the locally inertial rest frame of the emitting gas,
and we assume single transition event along the LoS.},
$1+z=\nu_{21}/\nu_{\rm obs}$ is the {\it relative redshift}\footnote{$z$ should not be confused with the cosmological redshift of the observer or that of the emitting gas in this paper.} between the emission and the ``observer'', 
and $T^{\rm obs}_{\rm CMB}(\hat n)$ is the CMB temperature along the LoS as seen by the ``observer''.
$\nu_{21}=1420$ MHz is the rest-frame frequency of the 21~cm transition.
$\tau_{\nu_{\rm obs}}$ is the 21~cm optical depth at the observed frequency.

In Eq.~(\ref{eq:deltatb}), the 21~cm optical depth is given by the LoS integration \citep[e.g.,][]{2007PhRvD..76h3005L},
\begin{equation}\label{eq:21opticaldepth}
	\tau_{\nu_{\rm obs}}=\int^{\rm obs}_{\rm em} \frac{3h^3A_{10}n^g_{\rm HI}T_{21}}{32\pi p^gT^g_{\rm s}}
	\phi(E^g-E_{21})\ud \lambda,
\end{equation}
where $E^g=p^gc$ is the energy of the photon.
The rest-frame 21~cm line profile $\phi(E^g-E_{21})$ is defined such that $\int \phi(E^g-E_{21})\ud E^g=1$. 
It can be approximated as a Dirac delta function, hence simplifying the 21~cm optical depth:
\begin{equation} \label{eq:21cmtau}
	\tau_{\nu_{\rm obs}}=\frac{3h^3cA_{10}T_{21}(1+z)}{32\pi E^2_{21}}
	\left(\frac{n^g_{\rm HI}}{T^g_{\rm s}}\left|\frac{\ud\lambda}{\ud z}\right|\right)\bigg|_{\rm em},
\end{equation}
where $n^g_{\rm HI}$ is the number density of the H~{\small I} gas,
$\lambda$ is the affine parameter along the ray, 
and the subscript `em' denotes the emission location where $E^g(\lambda)={E_{21}}$. 
The (inverse) LoS differential redshift, $|\ud\lambda/\ud z|$, 
is attributed to the gravitational acceleration 
and the Doppler effect from the motion of the medium.
It may diverge and thus cause $\tau_{\nu_{\rm obs}}$ to diverge, 
an extreme case due to the RSD component of the Doppler shift.
However, for actual line profiles with finite width,
the path integral in Eq.~(\ref{eq:21opticaldepth}) is always regular.
Moreover, unlike the number count in galaxy redshift surveys, 
the brightness temperature given by Eq.~(\ref{eq:deltatb}) is always finite 
even when $\tau_{\nu_{\rm obs}}$ diverges \citep{2012MNRAS.422..926M}.

The relationship between the gauge-invariant (GI) observed redshift 
and the gauge-dependent cosmological redshift is that, to linear order,
\begin{IEEEeqnarray}{rCl}\label{eq:redshift}
	& & (1+z)\frac{a(\eta_{\rm em})}{a(\eta_{\rm obs})} =  
	1+\hat n\cdot \left(\frac{\vec v_{\rm GI}}{c}\right)\bigg|^{{\rm obs}}_{{\rm em}} \nonumber \\
	& & + \left(\frac{\Psi_A}{c^2}+\frac{\Phi_H}{c^2}
	-\frac{\Phi}{c^2}+\frac{1}{3}\nabla^2 E\right)\bigg|^{{\rm obs}}_{{\rm em}} \nonumber \\
	& & -\int^{{\rm obs}}_{{\rm em}}\left(\frac{\dot\Psi_A}{c^2}+ \frac{\dot\Phi_H}{c^2}
	-\frac{c}{a}n^in^j\partial_i\dot V_j+\frac{1}{2}n^in^j\dot h_{ij}\right)\ud \eta \nonumber \\	& & =1+\frac{\delta z}{1+z}, 
\end{IEEEeqnarray}
where the overdot denotes the partial time derivative with respect to $\eta$,
the redshift variation $\delta z\equiv 1+z-a(\eta_{\rm obs})/a(\eta_{\rm em})$, 
and $\Psi_A$, $\Phi_H$, $V_j$, $h_{ij}$ are GI metric perturbations 
\citep{1980PhRvD..22.1882B, 2011PhRvD..84f3505B}.
The GI velocity perturbation $\vec v_{\rm GI}$ is evaluated for both the observer and the emitting gas.
It encodes the Doppler effect which sources the RSD.

In this paper, we adopt the conformal Newtonian gauge ($B=E=0$, $\Psi=\Psi_A$, $\Phi=\Phi_H$) 
and only considered scalar perturbations.
We define a GI time variable, the cosmic time $\bar\eta_{z_{\rm cos}}$ 
which in the background FLRW model corresponds to 
cosmological redshift $z_{\rm cos}$, $a(\bar\eta_{z_{\rm cos}})\equiv 1/(1+z_{\rm cos})$.
We also assume that the H~{\small I} gas is pressureless.
Thus, according to the conservation of momentum, $\dot{\vec v}+\mathscr{H}\vec v+\nabla\Psi=0$, 
where $\vec v$ is the scalar-mode peculiar velocity of the gas.

\subsection{The 21~cm brightness temperature in the optically-thin, post-heating, quasi-linear regime} \label{app:assumption}

As described in \S\ref{ssec:deltaTb}, the working assumptions of this paper are listed as:
$\tau_{\nu_{\rm obs}}\ll 1$, $T^g_{\rm s}(t_{\rm em})\gg (1+z)T^{\rm obs}_{\rm CMB}(\hat n)$
and $|\delta| \ll 1$. 
Particularly, combining the optically-thin condition with Eq.~(\ref{eq:21cmtau}) implies that 
$\ud z/\ud \lambda<0$ along the ray with respect to the emitting gas, 
so that $z(\lambda)$ is a monotonic function. As a result, the real-to-redshift-space coordinate conversion is always monotonic, i.e., no ``Finger-of-God'' effect for 21~cm radiation.

Plugging all the approximations and Eq.~(\ref{eq:21cmtau}) into Eq.~(\ref{eq:deltatb})
yields the 21~cm brightness temperature for a fixed observer, to linear order, 
\begin{IEEEeqnarray}{rCl}
	& & \delta T_{\rm b}(\eta_{\rm obs}, \vec x, \nu_{\rm obs}, \hat n)  \approx  \frac{3h^3cA_{10}T_{21}n^g_{\rm HI}|_{\rm em}}{32\pi E^2_{21}}
	\left|\frac{\ud\lambda}{\ud z}\right| \nonumber \\ 
	&  \approx & \frac{3c^3A_{10}T_{21}n^g_{\rm HI}|_{\rm em}}{32\pi\nu_{21}^3(1+z)}
	\left[H(\eta)\left(1-\frac{\Psi}{c^2}\right)-\frac{1}{a}\frac{\dot\Phi}{c^2} \right.\nonumber \\
	& & \left. + \frac{1}{a}\hat n\cdot(\hat n\cdot\nabla\vec v)\right]^{-1}\Bigg|_{\rm em}\nonumber\\
	& = & \frac{3c^3A_{10}T_{21} n^g_{\rm HI}|_{\rm em}}{32\pi\nu_{21}^3(1+z)H(\eta_{\rm em})} 
	\left(1-\frac{\Psi}{c^2}-\frac{1}{\mathscr{H}}\frac{\dot\Phi}{c^2} \right.\nonumber \\
	& & \left.+\frac{1}{\mathscr{H}}\frac{\partial v_\parallel}{\partial r}\right)^{-1}\Bigg|_{\rm em}\nonumber\\
	& \approx & \frac{3c^3A_{10}T_{21}\bar n_{\rm HI}(\bar\eta_{z_{\rm em}})}{32\pi\nu_{21}^3(1+z)H(\bar\eta_{z_{\rm em}})} \nonumber \\
	& & \times \frac{(1+\delta_{\rm HI}|_{\rm em})\left(1+\frac{\dot{\bar n}_{\rm HI}}{\bar n_{\rm HI}}(\bar\eta_{z_{\rm em}})\delta \eta_{\rm em}\right)}
	{1+\frac{\dot H}{H}(\bar\eta_{z_{\rm em}})\delta \eta_{\rm em}
	+\left(\frac{1}{\mathscr{H}}\frac{\partial v_\parallel}{\partial r}-\frac{1}{\mathscr{H}}\frac{\dot\Phi}{c^2}-\frac{\Psi}{c^2}\right)\Big|_{\rm em}}, \label{eq:tbrz}
\end{IEEEeqnarray}
where $n_{\rm HI}|_{\rm em}\equiv\bar n_{\rm HI}(\eta_{\rm em})(1+\delta_{\rm HI}|_{\rm em})$
defines the fluctuations of H~{\small I} distribution 
on the hypersurface of $\eta_{\rm em}$\footnote{Here 
instead of taking the H~{\scriptsize I} number density in the gas rest frame, 
we take the value measured in the perturbed FLRW frame. 
The difference is second-order in $|v|/c$ and thus negligible,
so we work with this approximation throughout the paper.},
and $\delta\eta_{\rm em}\equiv\eta_{\rm em}-\bar\eta_{z_{\rm em}}$,
$1+z_{\rm em}\equiv (1+z)(1+z_{\rm obs})=1/\langle a(\eta_{\rm em})\rangle$.
The corresponding variation in the cosmological redshift can be defined as 
$\delta z_{\rm em}\equiv 1+z_{\rm em}-1/a(\eta_{\rm em})$.
We can show that $\delta z_{\rm em}/(1+z_{\rm em})=\mathscr{H}(z_{\rm em})\delta\eta_{\rm em}$.
If the observer is on earth at the present, $z_{\rm obs}=0$, $z_{\rm em}=z$ and $\delta z_{\rm em}=\delta z$.
Note that $1+z_{\rm em}=\nu_{21}/\nu_{\rm obs,0}$, 
even when the emission event is not on the past light cone of the present-day earth observer.
Using Eq. (\ref{eq:redshift}), we obtain the expression for the redshift variation in the Newtonian gauge, 
\begin{IEEEeqnarray}{rCl}\label{eq:locationshift}
	& & \frac{\delta z_{\rm em}}{1+z_{\rm em}}  =  \frac{\delta z}{1+z}+\frac{\delta z_{\rm obs}}{1+z_{\rm obs}} \\
	& & =\left(\hat n\cdot\frac{\vec v}{c}+ \frac{\Psi}{c^2}\right)\bigg|^{{\rm 0}}_{{\rm em}} 
	 -\left(\int^{{\rm obs}}_{{\rm em}}+\int^{{\rm 0}}_{{\rm obs}}\right)\left(\frac{\dot \Psi}{c^2}+ \frac{\dot\Phi}{c^2}\right)\ud \eta, \nonumber\\
	& &  \delta z_{\rm obs}  \equiv  z_{\rm obs}-\tilde z_{\rm obs}, 
	 \qquad \langle\delta z_{\rm em}\rangle=\langle\delta z\rangle=\langle\delta z_{\rm obs}\rangle=0,\nonumber\\
	& &  \langle\delta\eta_{\rm em}\rangle= \langle\delta\eta_{\rm obs}\rangle=0.\nonumber
\end{IEEEeqnarray}

Eq.~(\ref{eq:tbrz}) is the full relativistic expression for the 21~cm brightness temperature
in the optically-thin, post-heating, quasi-linear limit, 
in agreement with Eq. (18) in \citet{2013PhRvD..87f4026H}.
The numerator and the denominator in the third line correspond to 
the redshift density perturbation \citep[$\delta_z$ in][]{2011PhRvD..84f3505B}
and the perturbation in the LoS length element extended by the gas per redshift bin ($|\ud\lambda/\ud z|$), 
respectively, both GI.
Note that the optically-thin limit is implied in the quasi-linear regime, 
since the RSD term (the velocity gradient) in the denominator is the only factor 
that can possibly make $\tau_{\nu_{\rm obs}}$ diverge.
Eq. (\ref{eq:tbrz}) also shows that $(1 + z)\delta T_{\rm b}$ 
is constant along the ray for fixed emission event and direction, 
as expected by the conservation of the photon distribution function during free streaming.


\subsection{The redshift-space expansion of the brightness temperature and peculiar velocity effects} \label{app:deltaTbz}

21~cm photons seen by a fixed observer at a fixed redshift (Eq. [\ref{eq:deltatb}])
are actually emitted from different distances for different directions on the observer's past light cone.
This is the so-called light-cone (LC) effect, 
mainly caused by the peculiar velocity of the gas.
The extra anisotropy from the LC effect poses a major computational challenge to 
directly using Eq. (\ref{eq:tbrz}) to obtain the observed 21~cm signal.
Also, it breaks the azimuthal symmetry around any mode vector $\vec k$ in harmonic analysis, 
different from the CMB case.
Besides the LC effect, H~{\small I} peculiar velocities give rise to
the RSD effect in the 21~cm intensity signal, as we see in Eq. (\ref{eq:tbrz}).
The RSD occurs in the LoS length element, $|\ud\lambda/\ud z|$, 
as part of the Jacobian between the real- and redshift-space coordinates  
\footnote{The 1D distortion along the LoS in the intensity mapping case
is thus distinct from the traditional RSD effect in galaxy surveys 
which occurs in the 3D volume element \citep{2013PhRvD..87f4026H}. }.
These effects need to be accounted for.

Our approximate solution is to rewrite the expression from the redshift-space point of view, 
by its Taylor expansion around the GI coordinate $(\bar\eta_{z_{\rm em}}, \vec x-s\hat n)$.
It is the ensemble-averaged location of emission, as described in \S\ref{ssec:deltaTb}.
The comoving radial distance in the redshift space reads
\begin{IEEEeqnarray}{rCl}
	s & \equiv & \int^{z_{\rm em}}_{z_{\rm obs}}\frac{c\ud\tilde z}{H(\tilde z)} 
	= c(\bar\eta_{z_{\rm obs}}-\bar\eta_{z_{\rm em}})
	= c\langle(\eta_{\rm obs}-\eta_{\rm em})\rangle \nonumber \\
	& = & r(\hat n,z)+\frac{c\delta z_{\rm em}}{H(z_{\rm em})}-\frac{c\delta z_{\rm obs}}{H(z_{\rm obs})}
	=\langle r(\hat n,z)\rangle.
\end{IEEEeqnarray}
Eq. (\ref{eq:locationshift}) shows that $|\delta z|/(1+z)\ll1$ for most of the emitting patches during the EoR,
so that the Taylor series up to linear order of $\delta\eta$ and $(s-r)$
should be a good approximation for the signal.
Discarding the negligible contribution from metric perturbations 
(the Sachs-Wolfe and integrated Sachs-Wolfe effect)
in Eq. (\ref{eq:locationshift}) yields
\begin{equation}
	\delta\eta_{\rm em}\approx \frac{1}{\mathscr{H}(z_{\rm em})}\frac{v_\parallel}{c}\bigg|^{\rm em}_0,
	\qquad s-r\approx \frac{v_\parallel}{\mathscr{H}}\bigg|^{\rm em}_{\rm obs}.
\end{equation}
Throughout the paper we have dropped the contributions from the observer's site
since they only affect the monopole and dipole of the observed anisotropy.
Thus, we obtain the following expression for $\Theta$ to linear order:
\begin{IEEEeqnarray}{rCl}\label{eq:thetalin}
	& & \Theta(\eta_{\rm obs}, \vec x, \nu_{\rm obs}, \hat n) \nonumber\\
	& = &   \frac{n_{\rm HI}|_{\rm em}/\bar n_{\rm HI}(z_{\rm em})}
	{1+\frac{\dot H}{H}( z_{\rm em})\delta\eta_{\rm em}
	+\left(\frac{1}{\mathscr{H}}\frac{\partial v_\parallel}{\partial r}-\frac{1}{\mathscr{H}}\frac{\dot\Phi}{c^2}-\frac{\Psi}{c^2}\right)\Big|_{\rm em}}-1\nonumber\\
	& \approx & \delta_{\rm HI}-\frac{1}{\mathscr{H}}\frac{\partial v_\parallel}{\partial r}\frac{n_{\rm HI}}{\bar n_{\rm HI}}-(s-r)\frac{\partial\delta_{\rm HI}}{\partial r}
	+\left(\frac{\dot n_{\rm HI}}{\bar n_{\rm HI}}-\frac{\dot H}{H}\frac{n_{\rm HI}}{\bar n_{\rm HI}}\right)\delta \eta_{\rm em} \nonumber \\
	& \approx & \delta_{\rm HI} - \frac{1}{\mathscr{H}}\frac{\partial v_\parallel}{\partial r}\frac{n_{\rm HI}}{\bar n_{\rm HI}}
	-\frac{\partial\delta_{\rm HI}}{\partial r}\frac{v_\parallel}{\mathscr{H}}\nonumber\\
	& & +\left[\frac{\dot n_{\rm HI}}{\bar n_{\rm HI}}
	-\left(\frac{\dot{\mathscr{H}}}{\mathscr{H}}-\mathscr{H}\right)\frac{n_{\rm HI}}{\bar n_{\rm HI}}\right]\frac{v_\parallel}{c\mathscr{H}},
\end{IEEEeqnarray}
where all variables on the right-hand side of the last line above are evaluated at $(\bar\eta_{z_{\rm em}}, \vec x-s\hat n)$.


The expression above can be simplified using the conservation of hydrogen number,
$\dot n_{\rm H}+\nabla\cdot(n_{\rm H}\vec v)+3(\mathscr{H}-\dot\Phi/c^2) n_{\rm H}=0$, 
combined with the fact that $n_{\rm HI}=n_{\rm H}x_{\rm HI}$. This leads to 
\begin{equation}
	\dot n_{\rm HI}+\nabla\cdot(n_{\rm HI}\vec v)+3(\mathscr{H}-\dot\Phi/c^2) n_{\rm HI}
	=n_{\rm H}(\dot x_{\rm HI}+\nabla\cdot(x_{\rm HI}\vec v)).
\end{equation}
Inserting it into Eq. (\ref{eq:thetalin}) yields
\begin{IEEEeqnarray}{lll}\label{eq:tbfinal}
	& & \Theta(\eta_{\rm obs}, \vec x, \nu_{\rm obs}, \hat n)  \approx  
	\delta_{\rm HI}-\frac{n_{\rm HI}}{\bar n_{\rm HI}}\frac{\partial_r v_\parallel}{\mathscr{H}}  \\
	& & +\left[\frac{\dot x_{\rm HI}}{\bar x_{\rm HI,m}}-\frac{c\partial_r n_{\rm HI}}{\bar n_{\rm HI}}
	-\left(\frac{\dot{\mathscr{H}}}{\mathscr{H}}+2\mathscr{H}\right)\frac{n_{\rm HI}}{\bar n_{\rm HI}}\right]\frac{v_\parallel}{c\mathscr{H}}, \nonumber
\end{IEEEeqnarray}
where the second term on the right-hand side accounts for the RSD
and the third term for the LC effect.

Figure~1 in \citet{2013PhRvD..87f4026H} shows that at low redshifts ($z\sim1-2$), 
the RSD is the dominant effect compared with the LC effect
and metric perturbation terms in the full expression (\ref{eq:tbrz}).
In this paper we have also neglected the LC and relativistic effects for the EoR signal, 
only keeping the RSD term, 
though the former may cause a change of the signal by a factor of order unity
\citep{2019MNRAS.490.1255C}.

\subsection{The resulting 21~cm brightness temperature}
\label{app:deltaTbdetail}

In Eq.~(\ref{eq:tbpostheating}), 
the dimensional factor $T_0(z)$, dependent on the relative redshift $z$ between the emitter and the observer, 
is defined as 
\begin{equation}
	T_0(z) \equiv \frac{3c^3A_{10}T_{21}\bar n_{\rm H}(z_{\rm em})}{32\pi\nu_{21}^3(1+z)H(z_{\rm em})},
\end{equation}
where $A_{10}=2.85\times 10^{-15}$ s$^{-1}$ is the 21~cm spontaneous emission rate,
$T_{21}\equiv h\nu_{21}/k_{\rm B}$,
and $\bar n_{\rm H}(z_{\rm em}) = \left(3\Omega_{\rm b} H_0^2X_{\rm P}/8\pi G m_{\rm H}\right)(1+ z_{\rm em})^3$, 
($X_{\rm P}$ is the cosmic hydrogen mass abundance and $m_{\rm H}$ is the mass of atomic hydrogen).
The global 21~cm brightness temperature (the ensemble-averaged value of the monopole) is then given by
\begin{eqnarray} \label{eq:Tbglobal}
	& & \bar{\delta T_{\rm b}}(\nu_{\rm obs})  =   T_0(z)\bar x_{\rm HI, m}(z_{\rm em})
	=\frac{3c^3A_{10}T_{21}\bar n_{\rm HI}(z_{\rm em})}{32\pi\nu_{21}^3(1+z)H(z_{\rm em})} \nonumber \\
	& & \approx   (27{\rm mK})\left(\frac{\Omega_{\rm b}h^2}{0.022}\right)
	\sqrt{\frac{0.14}{\Omega_{\rm m}h^2}\frac{1+z_{\rm em}}{10}}\bar x_{\rm HI, m}(z_{\rm em}).
\end{eqnarray}

As for the fluctuation part, we work in the Fourier space 
for the temperature anisotropy $\Theta$, defined in \S\ref{ssec:deltaTb}.
By definition, $\Theta(\eta_{\rm obs}, \vec x,\nu_{\rm obs}, \hat n)
=\delta T_{\rm b}(\eta_{\rm obs}, \vec x, \nu_{\rm obs}, \hat n)/\bar{\delta T_{\rm b}}(\nu_{\rm obs})-1$.
Taking into account the linear-theory relation 
$\vec v(\eta,\vec k)=i(\hat k/k)\mathscr{H}(\eta)\delta(\eta,\vec k)$
during the matter-dominated era \citep{2004MNRAS.352..142B}, 
we obtain from Eq.~(\ref{eq:tbpostheating}) that
\begin{IEEEeqnarray}{rll}
	& & \Theta  (\eta_{\rm obs}, \vec k, \nu_{\rm obs}, \hat n) \approx \delta_{\rm HI}(\vec k, z_{\rm em})e^{-i\mu ks} + \mu^2\delta(\vec k, z_{\rm em})e^{-i\mu ks} \nonumber\\
	& &  + \mathscr{C}_{\delta_{\rm HI}\delta}(\vec k, z_{\rm em}, \hat n)e^{-i\mu ks},
\end{IEEEeqnarray} 
where the cross-term is 
$\mathscr{C}_{\delta_{\rm HI}\delta}(\vec k, z_{\rm em},\hat n)  \equiv  
\int\frac{\ud^3\vec k'}{(2\pi)^3}\delta_{\rm HI}(\vec k-\vec k')\delta(\vec k')(\hat k'\cdot\hat n)^2$.
The last two terms on the RHS of the equation above are due to the RSD, 
carrying intrinsic angular dependence \citep{1997PhRvD..56..596H}.
Before and during the early phase of reionization, 
the convolution $\mathscr{C}_{\delta_{\rm HI}\delta}$ is a second-order term 
so that the azimuthal symmetry around $\vec k$ is preserved
\citep{2005ApJ...624L..65B, 2006ApJ...653..815M}.
However, once $\delta_{\rm HI}\sim 1$, 
the convolution term between the patchy H~{\small I} field and the RSD may not be negligible
even if the velocity field is still linear, 
breaking the azimuthal symmetry.

We leave the investigation of the convolution term to a future work and drop it in this paper. 
This assumption is valid at large scales where $\delta_{\rm HI} < 1$ still holds. 
Under this approximation, 
$\Theta(\eta_{\rm obs}, \vec k, \nu_{\rm obs}, \hat n)=\Theta(\eta_{\rm obs}, \vec k, \nu_{\rm obs}, \mu)$, 
i.e. its dependence on the LoS is only through functions of $\mu$.
This yields the final expression, Eq.~(\ref{eq:tbkfinal}).

\section{General expression for temperature and polarization angular power spectra}\label{app:generalCl}

In \S\ref{ssec:polarization} we derive the angular power spectra of 21~cm temperature and polarization anisotropies, 
based on the expression for the temperature fluctuations given by Eq.~(\ref{eq:tbkfinal}),
where $\Theta$ is apparently sourced by $\delta_{\rm HI}$, and $\delta$
(via the RSD correction).
However, as shown by Eq.~(\ref{eq:thetalin}), 
other physical effects, e.g., the LC effect and relativistic metric perturbations, might affect the 21~cm radiation. 
For interested readers, here we present expressions for the general case in which 
multiple cosmological fields source temperature fluctuations
and each source has its own transfer function.
Those source fields are require to be statistically homogeneous and isotropic,
and the temperature transfer functions should satisfy the azimuthal symmetry around $\hat k$.
In other words, $\Theta(\eta_{\rm obs}, \vec k, \nu_{\rm obs}, \mu)=\sum_i S_i(\vec k, z_{\rm em})T^{\rm T}_i(\eta_{\rm obs}, k, \nu_{\rm obs}, \mu)$, 
where $S_i$ represent the initial conditions for the $i$-th source and $T^{\rm T}_i$ the temperature transfer functions.

The (equal-time) auto-power spectra and cross-power spectra of the initial source fields are given by
$\langle S^*_i(\vec k)S_i(\vec k')\rangle=(2\pi)^3P_{i}(k)\delta^{(3)}_{\rm D}(\vec k-\vec k')$
and $\langle S^*_i(\vec k)S_j(\vec k')\rangle=(2\pi)^3P_{ij}(k)\delta^{(3)}_{\rm D}(\vec k-\vec k')$.
Also, for each source, we can define multipole moments of the temperature and polarization transfer functions,
\begin{widetext}
\begin{IEEEeqnarray}{rCl} 
	T^{\rm T}_{i,\ell}(\eta_{\rm obs}, k, \nu_{\rm obs}) & \equiv & 
	\frac{1}{(-i)^\ell}\int^1_{-1}\frac{\ud\mu}{2}\mathscr{P}_\ell(\mu)T^{\rm T}_i(\eta_{\rm obs}, k, \nu_{\rm obs}, \mu),\\
	\mathscr{T}^{\rm E}_{i,\ell}(\eta_{\rm obs}, k, \nu_{\rm obs}) & \equiv & \frac{3}{4}\sqrt{\frac{(\ell+2)!}{(\ell-2)!}}
	\int^{\bar\eta_{z_{\rm obs}}}_{\bar\eta_{z_{\rm em}}}g(\eta')T^{\rm T}_{i,2}(\eta', k, \nu')
	\frac{j_\ell[ck(\bar\eta_{z_{\rm obs}}-\eta')]}{[ck(\bar\eta_{z_{\rm obs}}-\eta')]^2}\ud \eta'.
\end{IEEEeqnarray}
Recall that $1+z_{\rm em}=(1+z)(1+z_{\rm obs})$ and $z$ is the observed (relative) redshift.

Hence, the angular power spectra of temperature and $E$-mode polarization anisotropies,
for a generic observer located at $(\eta_{\rm obs},\vec x)$ and observing at frequency $\nu_{\rm obs}$, are expressed as
\begin{IEEEeqnarray}{rCl} 
	C^{\rm EE}_\ell & = & \frac{2}{\pi}\int k^2\ud k
	\Big[\sum_iP_i(k, z_{\rm em})\left(\mathscr{T}^{\rm E}_{i,\ell}(\eta_{\rm obs}, k, \nu_{\rm obs}) \right)^2 
	+ \sum_{i\neq j}P_{ij}(k, z_{\rm em})\mathscr{T}^{\rm E}_{i,\ell}(\eta_{\rm obs}, k, \nu_{\rm obs}) 
	\mathscr{T}^{\rm E}_{j,\ell}(\eta_{\rm obs}, k, \nu_{\rm obs}) \Big],\\
	C^{\rm TT}_\ell & = & \frac{2}{\pi}\int k^2\ud k\Big[\sum_iP_i(k, z_{\rm em})\left(T_{i,\ell}(\eta_{\rm obs}, k, \nu_{\rm obs}) \right)^2 
	+ \sum_{i\neq j}P_{ij}(k,z_{\rm em})T_{i,\ell}(\eta_{\rm obs}, k, \nu_{\rm obs}) T_{j,\ell}(\eta_{\rm obs}, k, \nu_{\rm obs}) \Big].
\end{IEEEeqnarray}
The cross-power spectrum between the temperature and the $E$-mode polarization is 
\begin{equation}
	C^{\rm TE}_\ell = \frac{2}{\pi}\int k^2\ud k\Big[\sum_iP_i(k, z_{\rm em})T_{i,\ell}(\eta_{\rm obs}, k, \nu_{\rm obs}) 
	\mathscr{T}^{\rm E}_{i,\ell}(\eta_{\rm obs}, k, \nu_{\rm obs})
	+ \sum_{i\neq j}P_{ij}(k, z_{\rm em})T_{i,\ell}(\eta_{\rm obs}, k, \nu_{\rm obs})\mathscr{T}^{\rm E}_{j,\ell}(\eta_{\rm obs}, k, \nu_{\rm obs})\Big].
\end{equation}
\end{widetext}

\section{Previous modeling of the 21~cm polarization}\label{app:signalcompare}

Following Eq.~(\ref{eq:quadTb}) and the standard procedure, 
the formulae of 21~cm polarization angular power spectra in BL05 (their Eqs.~13 and 25) can be rewritten using {\it our} notation\footnote{Note that in BL05, the multipole moments of the transfer functions due to the RSD, $T^{\rm T}_{\delta,\ell}$ and $\mathscr{T}^{\rm E}_{\delta,\ell}$, take different forms from ours. This can be understood by comparing Eq.~(\ref{eq:Thetaell}) with Eq.~(\ref{eq:quadTb}).} as  
\begin{widetext}
\begin{IEEEeqnarray}{rCl}
    C^{\rm EE}_\ell & = & \frac{2}{\pi}\int k^2\ud k
	\Big\{  \left[{\bar x_{\rm HI}}^{-2}\,P_x(k)+P_\delta(k)\right]\left(\mathscr{T}^{\rm E}_{{\rm HI},\ell}(k) \right)^2 + \, P_\delta(k)\left(\mathscr{T}^{\rm E}_{\delta,\ell}(k) \right)^2 
	+ \, 2\,P_\delta(k)\,\mathscr{T}^{\rm E}_{{\rm HI},\ell}(k)\,\mathscr{T}^{\rm E}_{\delta,\ell}(k)	\Big\},\qquad\qquad \label{eqn:BL05EE}\\
	C^{\rm TE}_\ell & = & \frac{2}{\pi}\int k^2\ud k
	\Big\{  \left[{\bar x_{\rm HI}}^{-2}\,P_x(k)+\,P_\delta(k)\right]T^{\rm T}_{{\rm HI},\ell}(k) \mathscr{T}^{\rm E}_{{\rm HI},\ell}(k) \nonumber\\
	& & + P_\delta(k)\,T^{\rm T}_{\delta,\ell}(k) \mathscr{T}^{\rm E}_{\delta,\ell}(k)  + P_\delta(k)\,\big[T^{\rm T}_{{\rm HI},\ell}(k)\mathscr{T}^{\rm E}_{\delta,\ell}(k)
	+T^{\rm T}_{\delta,\ell}(k)\mathscr{T}^{\rm E}_{{\rm HI},\ell}(k)\big]\Big\}, \label{eqn:BL05TE}
\end{IEEEeqnarray}
\end{widetext}
where, again, all the transfer functions are evaluated at $(\eta_0,\nu_{\rm obs,0})$,
all the power spectra and $\bar x_{\rm HI}$ are evaluated at $z_{\rm em}$, 
and the dimensional results should be obtained by multiplying the extra factor of $\left[T_0(z_{\rm em})\bar x_{\rm HI,m}(z_{\rm em})\right]^2$.

Comparing Eqs.~(\ref{eqn:BL05EE}) and (\ref{eqn:BL05TE}) with Eqs.~(\ref{eq:EEps}) and (\ref{eq:TEps}), it is realized that the approximations made in BL05 are $P_{\rm HI}(k)\approx {\bar x_{\rm HI}}^{-2}\,P_x(k)+\,P_\delta(k) $ and 
$P_{{\rm HI}\,\delta}(k) \approx P_\delta(k)$. Since $\delta_{\rm HI} = \delta x_{\rm HI} + \delta + \delta x_{\rm HI}\, \delta $, we have $P_{\rm HI}(k) =  {\bar x_{\rm HI}}^{-2}\,P_x(k)+\,P_\delta(k) + 2 {\bar x_{\rm HI}}^{-1}\,P_{x_{\rm HI}\,\delta} \,+ $ (higher-order cross terms), and  $P_{{\rm HI}\,\delta}(k) = P_\delta(k) + {\bar x_{\rm HI}}^{-1}\,P_{x_{\rm HI}\,\delta} \,+ $ (higher-order cross terms). Therefore, not only did BL05 ignore the higher-order cross terms due to $\delta x_{\rm HI}\, \delta $, but also it ignored the leading-order cross term $P_{x_{\rm HI}\,\delta}$. 
In comparison, our formalism is expanded on the H~{\small I} density fluctuations, 
which automatically corrects for both issues.


\bibliography{21cm_cosmology}{}
\bibliographystyle{aasjournal}



\end{document}